\documentclass[twocolumn]{svjour3}                     
\smartqed  
\usepackage{graphicx}
\usepackage{mathptmx}      
\usepackage{natbib}
\journalname{SSRv}
\begin{document}

\title{Properties and selected implications of 
magnetic turbulence for interstellar medium, Local Bubble and solar wind 
}

\newcommand{\be}{\begin{equation}}
\newcommand{\ee}{\end{equation}}
\newcommand{\ba}{\begin{eqnarray}}
\newcommand{\ea}{\end{eqnarray}}
\newcommand{\eps}{\varepsilon}


\author{A. Lazarian, A. Beresnyak, H. Yan, M. Opher \& Y. Liu}

\authorrunning{Lazarian et al.} 

\institute{A. Lazarian, A. Beresnyak \at
              Department of Astronomy, University of Wisconsin-Madison \\
           M. Opher \at
           Department of Physics and Astronomy, George Mason University\\
           H. Yan \at
           CITA, University of Toronto, Canada\\
           Liu, Y. \at
          Univeristy of New Hampshire 
}


\maketitle

\begin{abstract}
Astrophysical fluids, including interstellar and interplanetary medium, are magnetized and turbulent. Their appearance, evolution, and overall properties are determined by the magnetic turbulence that stirs it. We argue that examining magnetic turbulence at a fundamental level is vital to understanding many processes. A point that frequently escapes the attention of researchers is that magnetic turbulence cannot be
confidently understood only using ``brute force'' numerical approaches. In this review we illustrate this point on a number of examples,
including intermittent heating of plasma by turbulence, interactions of turbulence with cosmic rays and effects of turbulence on the rate of magnetic reconnection. We show that the properties of magnetic turbulence may vary considerably in various environments, e.g. imbalanced turbulence in solar wind differs from balanced turbulence and both of these differ from turbulence in partially ionized gas.  Appealing for the necessity of more observational data on magnetic fields, we discuss a possibility of studying interplanetary turbulence using alignment of Sodium atoms in the tail of comets.
\end{abstract}

\section{Introduction}
\label{intro}
It is well known that astrophysical
fluids are magnetized and turbulent (see Armstrong \& Woo 1981, Armstrong et al. 1995,  Verdini 
\& Velli 2007 and ref. therein). Turbulence is known to affect most of the properties of fluids, e.g.
thermal conductivity, propagation of waves and energetic particles, magnetic field generation etc. (see Kazantsev 1968,
Moffatt 1978, Dmitruk et al. 2001, Schlickeiser 2003, Vishniac, Lazarian \& Cho 2003, Cranmer \& van Ballegooijen 2005 and references therein). Local Bubble, solar wind and interstellar medium are not exceptions: the fluids there are turbulent.
A substantial progress in understanding of the media above has been achieved
using numerical simulations. It is very encouraging that present codes can produce simulations that resemble observations. However, one may wonder to  what extent numerical results reflect reality. The answer to this question is not as simple as it may look.

It is easy to justify a cautious approach to
the interpretation of the numerical simulations. Indeed, a meaningful numerical
representation of the turbulent fluid requires some basic non-dimensional
combinations of the physical parameters of the simulation to be
similar to those of the real ISM. One such is the ``Reynolds number'',
$Re$, the ratio of the eddy turnover time of a parcel of gas to the
time required for viscous forces to slow it appreciably. A similar
parameter, the ``magnetic Reynolds number'', $Rm$, is the ratio of the
magnetic field decay time to the eddy turnover time. The properties of
flows on all scales depend on $Re$ and $Rm$. It is not realistic to
expect that we can in any foreseeable future simulate turbulent flows with
 $Re>10^8$ and $Rm>10^{16}$. Note that 
3D simulations for 512 cubes can have $Re$ and $Rm$ up to $\sim 2000$
and are limited by their grid sizes. 
It should be kept in mind that while low-resolution
observations show true large-scale features, low-resolution numerics may
potentially produce an incorrect physical picture.

How feasible is it, then, to strive to understand the complex
microphysics of astrophysical MHD turbulence? Substantial progress in
this direction is possible by means of ``{\it scaling laws}'', or analytical
relations between non-dimensional combinations of physical quantities
that allow a prediction of the motions over a wide range of $Re$ and $Rm$.
Even with its limited resolution, numerics is a great tool to {\it test} scaling laws.

On a basis of several selected examples we show that a "brute force" numerical approach has limited applicability while dealing with
interstellar medium, Local Bubble and solar wind. Instead we claim that to create an appropriate "tool box" that can be incorporated into simulations one must understand better underlying fundamental physical processes. Moreover, we can define the circumstances when one should and should not rely on the present-day numerical simulations only by better understanding these processes. For instance,
if magnetic reconnection is slow in collisional astrophysical environments, numerical simulations {\it cannot}
represent fluids there. We argue below that in turbulent fluids reconnection is fast and
therefore numerical simulations are not very different from the reality in this respect.   We believe that the solid progress in the studies of complex processes in magnetized astrophysical fluids must be based
on verification of theoretical constructions with observations. For this purpose we discuss a new  technique of
studying interplanetary magnetic fields using a subtle phenomenon of atomic alignment.   

We realize that within a relatively short review, one cannot cover in depth the properties and numerous implications of astrophysical turbulence.
Therefore while covering a few relevant subjects within the area our expertise,  we provide, wherever appropriate, references to extensive reviews   
and monographs. Although our review is intended for the special volume of the proceedings on the heliosphere and the Local Bubble, a number of examples in the review are from the area of interstellar medium. In fact, we feel that the existence of the gap  between the
interstellar and the heliospheric/interplanetary communities is not healthy, as the fundamental underlying physics that researchers deal with in these two cases is very similar. Moreover, we believe that with the Voyager spacecrafts approaching the interstellar medium (see Stone et al 2008, http://voyager.jpl.nasa.gov/) , the cross-polination between the fields should be enhanced.  

In what follows we discuss the spectrum and anisotropy of magnetic turbulence in different regimes, including the regime of balanced turbulence, imbalanced turbulence and viscosity-damped turbulence (\S 2). We consider turbulence intermittency  in \S 3 and the modification of  the magnetic turbulence in the presence of cosmic rays in \S 4. \S 5 is devoted to the selected implications of turbulence in interstellar and interplanetary medium, including scattering of cosmic rays, turbulent reconnection and perpendicular
diffusion of cosmic rays and heat.  We present an example of interplanetary turbulence studies using observations of the sodium tail of a comet in \S 6. Our main results are summarized in \S 7.

\section{Spectrum of turbulence and its anisotropy}

\subsection{Anisotropic MHD turbulence}

Magnetized turbulence is a tough and complex problem with many excellent monographs and reviews devoted to different aspect of it (see Biskamp 2003 and references therein). A broad outlook on the astrophysical implications of the turbulence can be found in a review by
Elmegreen \& Scalo (2004), while the effects of turbulence on molecular clouds and star formation are reviewed in McKee \& Ostriker (2007) and
Balesteros-Paredes et al. (2006). However, the issues of turbulence spectrum and its anisotropies, we feel, are frequently given less attention than  they deserve.  

 While turbulence is an extremely complex chaotic non-linear phenomenon, it
allows for a remarkably simple statistical description (see Biskamp 2003). If the injections
and sinks of the energy are correctly identified, we can describe
turbulence for {\it arbitrary} $Re$ and $Rm$.  The simplest
description of the complex spatial variations of any physical
variable, $X({\bf r})$, is related to the amount of change of $X$
between points separated by a chosen displacement ${\bf l}$, averaged
over the entire volume of interest. Usually the result is given in
terms of the Fourier transform of this average, with the displacement
${\bf l}$ being replaced by the wavenumber ${\bf k}$ parallel to ${\bf l}$
and $|{\bf k}|=1/|{\bf l}|$. For example, for isotropic turbulence the kinetic
energy spectrum, $E(k)dk$, characterizes how much energy resides at
the interval $k, k+dk$.  At some large scale $L$ (i.e., small $k$),
one expects to observe features reflecting energy injection. At small
scales, energy dissipation should be seen.  Between these two scales
we expect to see a self-similar power-law scaling reflecting the
process of non-linear energy transfer.

Thus, in spite of its complexity, the turbulent
cascade is self-similar over its inertial range. The physical variables are
proportional to simple powers of the eddy sizes over a large range of
sizes, leading to scaling laws expressing the dependence of certain
non-dimensional combinations of physical variables on the eddy
size. Robust scaling relations can predict turbulent properties on the
whole range of scales, including those that no large-scale numerical
simulation can hope to resolve. These scaling relations are extremely
important for obtaining an insight of processes on the small scales.

The presence of a magnetic field makes MHD turbulence anisotropic (Montgomery \& Turner 1981, Matthaeus et al. 1983, 
Shebalin et al. 1983, Higdon 1984, Goldreich \& Sridhar 1995, see Oughton, Dmitruk \& Matthaeus 2003 for a review). The
relative importance of hydrodynamic and magnetic forces changes with
scale, so the anisotropy of MHD turbulence does too.
Many astrophysical results, e.g. the dynamics of dust, scattering and acceleration of energetic particles, 
thermal conduction, can be obtained if the turbulence spectrum and its
anisotropy are known. As we discuss below, additional important insight can be obtained if 
we know turbulence intermittency. 

Estimates of turbulence anisotropy obtained in relation to the observations
of magnetic fluctuation of the outer heliosphere and solar wind  (see Zank \& Matthaeus 1992 and references therein)
provided, for an extended period of time, the only guidance for theoretical advances. This resulted in a picture of MHD
turbulence consisting of 2D "reduced MHD" perturbations carrying approximately 80\% of energy and "slab" Alfvenic waves
carrying the remaining 20\% of energy (see Matthaeus 2002 and references therein). In other words, in the suggested picture
the MHD turbulence was presented by two anisotropic components, one having wave vectors mostly perpendicular to magnetic
field (the 2D one), the other having them mostly parallel to magnetic field (the slab one). This model became a default one for
many calculations of the propagation of cosmic rays (see Bieber, Smith \& Matthaeus 1988, 
Bieber, Matthaeus \& Smith 1994). On the contrary, guided mostly
by compressible MHD numerical simulations, the interstellar community adopted a model of the MHD turbulence where
the basic MHD modes, i.e. slow, fast and Alfven are well coupled together and efficiently dissipate energy in shocks (Stone et  al. 1998,
Mac Low 1999). Little cross-talk between the two communities did not stimulate the interdisciplinary debates on the nature of MHD turbulence, which was regretful, as the magnetospheric community has the advantage of the in-situ spacecraft measurements.   

In spite of the intrinsic limitations of the "brute force" approach, 
we feel that reliable results can be obtained numerically if
the studies are  focused on a particular property of turbulence in
order to get a clear picture of the underlying physics occurring on
small scales (``microphysics'') that cannot be resolved in ``global''
interstellar simulations\footnote{By contrast, numerical
simulations that deal with many physical conditions simultaneously
cannot distinguish between the effects of different
processes. Moreover, they inevitably have a more restricted interval
of scales on which energy is injected by numerics, initial conditions,
or boundary conditions. Their results are, therefore, difficult to
interpret in physical terms.}.

We feel that it were the "focused" numerical simulations that allowed the validation of the Goldreich \& Shridhar (1995, henceforth GS95) model
of MHD turbulence.  Indeed, GS95 made predictions regarding relative motions parallel and perpendicular to {\bf B} for Alfvenic turbulence. 
The model did not predict the generation of any "slab" modes and, instead of pure 2D Alfvenic modes, predicted that most of the Alfvenic energy is
concentrated in the modes with a so-called "critical balance" between the parallel and perpendicular motions. The latter can be understood within intuitive picture where eddies mixing magnetic field perpendicular to its local\footnote{The notion of the direction being local is critical. Small eddies are affected by magnetic field in their vicinity, rather than a global field. No universal scalings are possible to obtain in the frame of  the mean magnetic field.} direction induce Alfvenic waves with the period equal to the period of the eddy rotation. This results in the scale-dependent anisotropy of velocity and magnetic perturbations, with the anisotropy being larger for smaller eddies.    

The relations predicted in GS95
were confirmed numerically for incompressible\footnote{As in any developing field, ongoing controversies
and competing ideas exist on how to improve the GS95 scalings (see Boldyrev 2005, 2006, Beresnyak \& Lazarian 2006, Gogoberidze 2007).} (Cho \& Vishniac 2000, Maron \& Goldreich 2001, Cho, Lazarian \& Vishniac
2002a) and compressible MHD turbulence\footnote{Some studies of MHD compressible turbulence, e.g. Vestuto et al. 2003 did not perform a decomposition of MHD perturbations into Alfven, slow and fast modes as it is done in CL02 and CL03. They did not use local system of reference for which the GS95 scaling is formulated. Therefore a direct comparison of their results with the GS95 predictions is difficult.} (Cho \& Lazarian 2002, 2003, henceforth CL02 and CL03). (see Cho, Lazarian \& Vishniac 2003 for a review). They are in good agreement with observed and inferred astrophysical spectra. A
remarkable fact revealed in Cho, Lazarian \& Vishniac (2002a) is that
fluid motions perpendicular to {\bf B} are identical to hydrodynamic
motions. This provides an essential physical insight and explains why
in some respects MHD turbulence and hydrodynamic turbulence are
similar, while in other respects they are different.

GS95 provided theoretical arguments in favor of low coupling between fast and Alfven modes, and low impact of slow modes to Alfven modes (see also Lithwick \& Goldreich 2001). This challenged the paradigm accepted by the interstellar community. While the decomposition of MHD perturbations into fundamental MHD waves was widely used in the literature (see Dobrowolny et al. 1980) it was usually assumed that the Alfvenic waves exist and interact with other waves for many periods (see a discussion in Zweibel, Hietsch \& Fan 2003). This is not the case of the GS95 model of turbulence, where the Alfven modes non-linearly decay within one wave period. This reduces the time of interaction and therefore the coupling. Interestingly enough, in GS95 model, the Alfvenic modes can affect slow modes, but the opposite is not true. These results were successfully tested in CL02 and CL03.   

Insight into fundamental properties of interstellar
turbulence has already paid dividends. For instance, theories of
cosmic ray propagation and dynamics of interstellar grains have been
 revised in view of better understanding of MHD turbulence\footnote{Incidentally, the concept of the scale-dependent anisotropy in respect to
 the {\it local} direction of magnetic field is very important. This local magnetic field is what actually affects energetic particles.} (see 
Cho \& Lazarian 2005 and references therein). Similarly, the issue of
turbulent support of molecular clouds requires revisions.  Indeed,
a widely
accepted view that the rapid decay of turbulence is caused by the
coupling of compressible and incompressible motions was disproved  with numerical simulations in
CL02. In fact, it was shown there that the rate of turbulent decay was shown to depend on the ways that
the turbulence is driven (Cho, Lazarian \& Vishniac 2003a). Moreover, contrary
to common beliefs, MHD motions do not die away at the scale at which
ambipolar diffusion damps hydrodynamic motions. As we discuss in \S 2.1 a 
new regime of viscosity-damped MHD turbulence emerges (Cho, Lazarian \& Vishniac 2002b).
 
\begin{figure*}[h!t] 
\includegraphics[width=6.5 in]{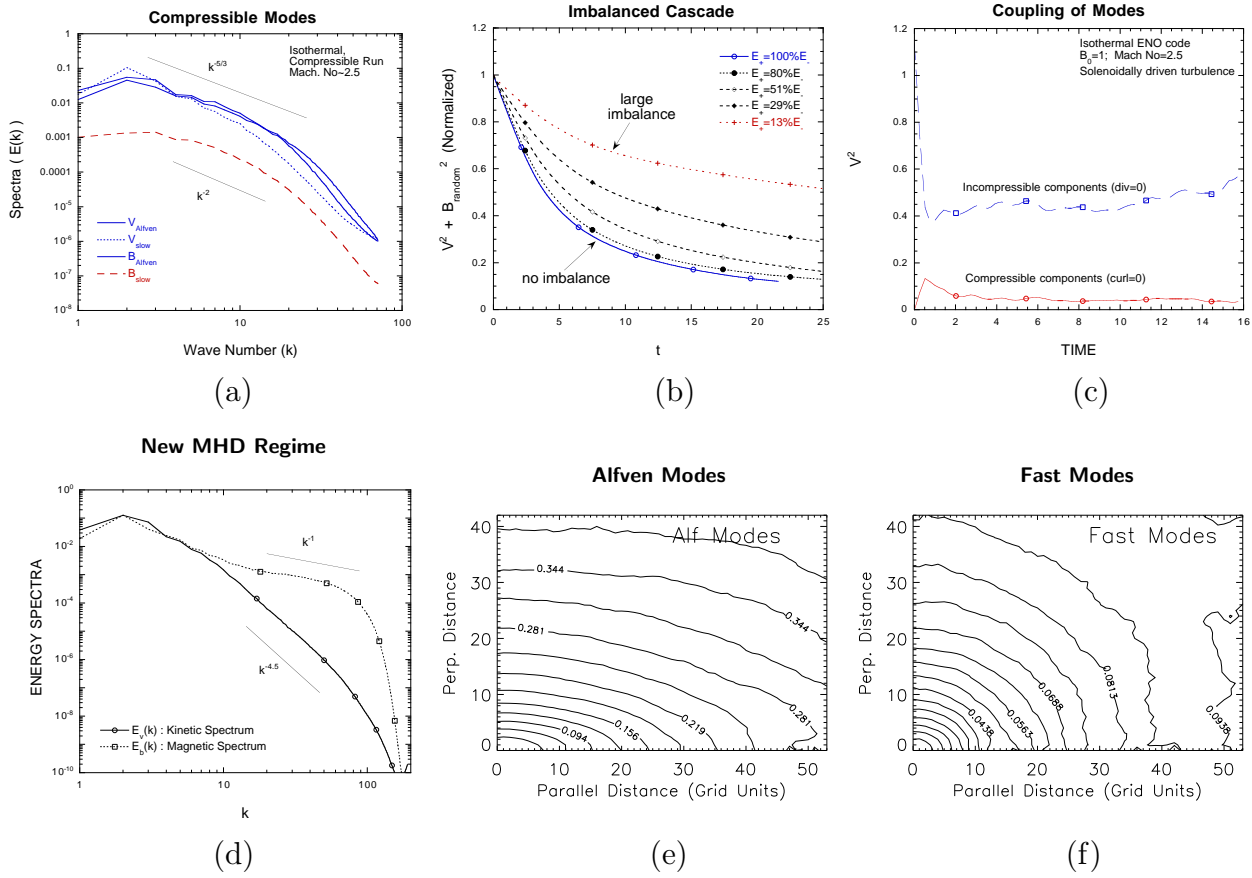}
\caption{{\it Statistics of MHD Turbulence:}  (a). Scaling of compressible motions for plasma
with magnetic pressure ten times the gas pressure. This regime is
important for molecular clouds (from CL02). The velocity (solid line)
and {\bf B} (dashed line) spectra are plotted against $k$ ($\equiv$
1/(eddy size)).  They show well defined statistical properties that
allow further fruitful applications.  (b). The simulations of decaying
turbulence show that the rate of the decay of the total energy is a
strong function of the imbalance of the energy contained in waves
initially moving in opposite directions. In the lowest curve, the
waves have the same amplitude. The energy decays more slowly with
imbalance, affecting how long turbulence can support molecular clouds
and whether transfer of energy between clouds is efficient (Cho,
Lazarian \& Vishniac 2002b). (c). The evolution of kinetic energy in
3-D compressible turbulence, initially started with
$\nabla\cdot v=0$ as if the gas were incompressible. The dashed curve
shows the evolution of the $\nabla\cdot v=0$ motions. We see that
Alfvenic turbulence creates only a marginal amount of compressible
motions, suggesting that Alfvenic modes should evolve independently of
the compressible cascade (CL02).  (d). Magnetic fluctuations persist
beyond the turbulent damping scale at large $k$, while hydrodynamic
fluctuations damp out in partially ionized gas (Lazarian, Vishniac \& Cho 2004).  This viscosity-dominated regime of turbulence may dominate small
scale structure of partially ionized gas. (e-f). Isocontours of equal
correlation for Alfven and fast modes (CL02). (e) The Alfvenic motions are
much more correlated along {\bf B} than perpendicular
to it. (f) In contrast, fast magnetosonic fluctuations show
essentially circular (isotropic) isocontours of correlation.}
\label{spectrum}
\end{figure*}

Some of the relevant results are illustrated in Figure\ref{spectrum}
 Contrary to the common expectation, the modes
exhibited nice scaling laws that allow further analytical and
numerical applications. 
For instance, numerical studies in CL02 and in CL03 reveal that the Goldreich-Shridhar scalings are
valid for the Alfvenic part of the turbulence cascade even in the
highly compressible regime (see also Beresnyak \& Lazarian 2006).

The density spectrum of MHD turbulence was studied in Beresnyak, Lazarian \& Cho (2006) and Kowal, Lazarian \& Beresnyak (2007). The spectrum was shown to be approximately Kolmogorov for low Mach numbers, and get flatter as the Mach number increases. The compression of gas by shocks was identified as the reason for the spectrum flattening. At the same time, it was shown that the logarithms of density have a Kolmogorov spectrum and GS95 anisotropies.

The statistical decomposition of MHD turbulence into Alfven, slow and fast modes suggested in CL02 was successfully tested for slow modes of magnetically dominated plasmas in CL03. The statistical procedure was further improved in 
Kowal \& Lazarian (2006), where wavelets rather than the Fourier transforms were used. There  the study of
spectra and anisotropy of velocity, magnetic field and density of compressible MHD turbulence was performed and the
advantages of the wavelets in comparison with the Fourier technique were revealed for studying turbulence with weak mean
field. For turbulence with $B_{mean}\sim \delta B$ the obtained results are consistent with the CL02 and CL03 studies. Nevertheless, it would be wrong to say that
we have a complete understanding of the scaling of MHD modes and their interactions. First of all, one should
distinguish weak and strong Alfvenic turbulence. The weak turbulence is essentially 2D, with the turbulent
cascade creating more structure perpendicular to magnetic field as the turbulence cascades (see Gaultier et al. 2000).  Such a cascade emerges 
when the driving of turbulence at the outer scale is weak, i.e. the injection velocity is much less than the Alfven velocity. 
Although the weak turbulence picture corresponds to the early representation of MHD turbulence (see discussion in \S 2.1), one should keep in mind that the strength of Alfvenic interactions increases with the decrease of the scale along the cascade. Therefore the Alfvenic turbulence 
gets eventually strong, while both the inertial range and the astrophysical utility of the weak Alfvenic cascade are limited. The interaction of weak Alfvenic
turbulence with fast modes has dependences on the angle between ${\bf B}$ and the wavevector, as was shown by Chandran (2005). In addition, we discuss in \S 4
that the fact that no slab waves appear in MHD simulations does not mean that such modes do not appear in realistic astrophysical
circumstances, e.g. when cosmic rays are present. Moreover, we show in \S 5 that in many circumstances, the way cascade is initiated and dissipates affect both the properties and implications of magnetic turbulence.

\subsection{Imbalanced Turbulence}

MHD turbulence in the presence of sources and sinks gets {\it imbalanced}, in the sense that
the flow of energy in one direction is larger than the flow of energy in the opposite direction.
Solar wind presents a vivid example of imbalanced turbulence with most waves near the Sun moving
in the direction away from the Sun.

While theories of balanced MHD turbulence enjoyed much attention, the theory of imbalanced turbulence\footnote{Another name for imbalanced turbulence is a turbulence with non-zero cross-helicity.} attracted less work, unfortunately (see Biskamp 2003 and references therein).  The
analytical results were obtained for {\it weak imbalanced turbulence}
(Galtier et al. 2002, Lithwick \& Goldreich 2003) and they are
applicable in a rather narrow range of imbalance ratios. Some earlier simulations of strong imbalanced turbulence
were limited to rather idealized set ups (Maron \& Goldreich 2001, Cho, Lazarian \& Vishniac 2002a), i.e.
for the initial state the results of the simulations of strong balanced turbulence were used, but the amplitudes
of waves moving in one direction were reduced. This did not allow making definitive conclusions about properties
of imbalanced turbulence.

We think that the best experimental data on the imbalanced regime is currently available from observations
of solar wind turbulence (e.g., Horbury 1999). This data,
collected by spacecrafts, is consistent with Kolmogorov $-5/3$ spectrum,
but does not provide sufficient insight
into the anisotropy with respect to the local magnetic field.
The imbalanced turbulence is not a rare exception, on the contrary,
such processes as preferential decay of a weaker wave and the escape
of turbulent energy from regions that generate perturbations make imbalanced
turbulence widely spread in various astrophysical circumstances.

Attempts to construct the model of stationary {\it strong} imbalanced
turbulence were done in Lithwick, Goldreich \& Sridhar (2007), Beresnyak \& Lazarian (2008b), Chandran (2008). 
Below we discuss only the model in Beresnyak \& Lazarian (2008b) as this is the only model
that agrees with numerical simulations performed so far. In view of a big picture, this model can be viewed as an 
 extention of GS95 model into the imbalanced regime. 
\begin{figure*}
\includegraphics{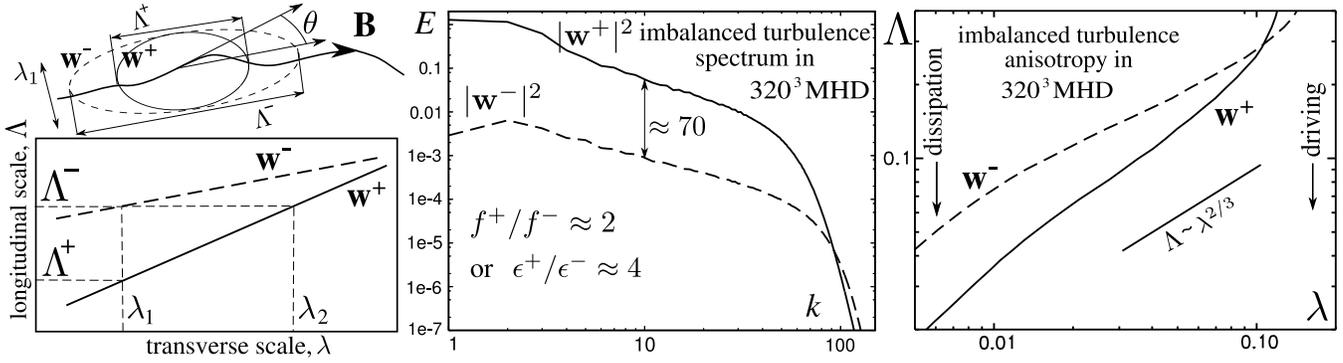}
\caption{Left upper: ${\bf w}^+$ wavepacket, produced
by cascading by ${\bf w}^-$ wavepacket is aligned with respect
to ${\bf w}^-$ wavepacket, but disaligned with respect
to the local mean field on scale $\lambda_1$, by the angle $\theta$.
Left lower: the longitudinal scale $\L$ of the wavepackets,
as a function of their transverse scale, $\l$;
Middle: the power spectrum of
energies for both waves in an imbalanced forced
incompressive $320^3$ numerical simulation. 
Right: the $\L(\l)$ dependence in the same simulation,
the length scales are in the units of the cube size.}
\label{imbalance}
\end{figure*}

While the classic formulation of the GS95 critical balance, based on causality, is unable
to describe consistently the imbalanced case, Beresnyak \& Lazarian (2008b) proposed a new way to 
introduce the balance between parallel and perpendicular modes. The new conditions for the critical balanced 
between the parallel and perpendicular modes of oppositely moving waves was obtained appealing to the process 
termed "propagation cascading". In the case of the balanced turbulence the "old" and "new" critical balance condition
results in the same GS95 scaling. However, in the case of imbalanced turbulence the
new formulation actually predicts {\it smaller} anisotropy for the stronger wave,
which directly contradicts old causal critical balance, but is consistent with
simulations.

\def\L{{\Lambda}}
\def\l{{\lambda}}

We assume that waves have different anisotropies, i.e. the dependence
of longitudinal scale $\L$ to transverse scale $\l$ is different for each kind of wave.
This situation is presented in Fig. 1, where some arbitrary
longitudinal scale $\L^-$ corresponds to the {\it two different} transverse scales,
$\l_1$ for weak wave $w^-$ and $\l_2$ for strong wave $w^+$.
$\L^+$ is a longitudinal scale of $w^+$ wave having
transverse scale $\l_1$. In case of strong turbulence,
we expect that at least the $w^-$ is being strongly
cascaded by $w^+$. In this case the most effective mixing
of $w^-$ on scale $\l_1$ will be obtained through $w^+$ motions
that are on the same scale.
The longitudinal scale for $w^-$ will be provided
by causal critical balance, since
its cascading is fast.

The cascading of $w^+$ is somewhat more complicated. Since
the amplitude of $w^-$ is not large enough
to provide strong perturbations in $w^+$, the $w^+$ will be perturbed weakly, and
the cascading timescale will be diminished according to the ``strength''
of the $w^-$, just like it does in weak turbulence. Moreover, now the $w^-$ eddies
will be cascading $w^+$ eddies with similar longitudinal scales,
which is the generic feature of weak cascading.

The perturbations provided by $w^-$ will have a transverse scale of $\l_1$.
In other words, the energy of $w^+$ will be transferred between $\l_2$ and $\l_1$.
The longitudinal scale for {\it cascaded} $w^+$ will be determined by
the propagation critical balance in the following way.
The wavepackets of $w^-$ are strongly aligned to the mean field
on scale $\l_1$ and therefore they are randomly oriented with respect
to the mean field at a larger scale $\l_2$.
The RMS angle of wavevector of $w^-$ eddies with respect to mean field on $\l_2$ will be around
$\theta\approx\delta b^+(\l_2)/v_A$. This slant of $w^-$ wavepackets will determine
the increase of $k_\|$ for newly cascaded $w^+$ packets at $\l_1$ (see Figure\ref{imbalance}).
The new interpretation
of critical balance in the strongly imbalanced case is that the
$k_\|$ of the weak wave increases due to the {\it finite lifetime}
of the wave packet, while in the strong wave it increases
due to the field wandering of the strong wave itself
{\it on larger scales}. This effect does not contradict the exact
MHD solution of the wave propagating in one direction, because
it requires the oppositely propagating wave as an intermediary.
The de-alignment of the cascaded strong wave is possible because
the weak wave, acting as a cascading agent, is strongly aligned
with the field lines on scale which is {\it different} (smaller)
than the scale of the strong wave it is acting upon.

While the model in Beresnyak \& Lazarian (2008b) is consistent with numerical simulations, higher
resolution simulations are definitely required. Moreover, the three models of strong imbalanced
turbulence, namely, Lithwick et al. (2007), Beresnyak \& Lazarian (2008b) and Chandran (2008), provide 
predictions to be compared with the solar wind observations. Naturally, additional processes should be accounted
for in the stellar wind studies, e.g. 
parametric instabilities, reflections from preexisting density fluctuations (see Leroy 1980,
Roberts et al. 1987, Hobery 1999,
Del Zanna et al. 2001). This calls for further studies of  imbalanced turbulence in compressible magnetized fluids.

\subsection{Viscous Turbulence}

When the magnetic turbulence takes place in viscous, but well conducting gas, its properties differ from
those described above. The effect of the neutral gas can act in a number respects as the proxy of
fluid viscosity\footnote{We, however, warn our reader not to identify the viscosity and the effects of neutral gas
on turbulence. The actual physics of neutral-ion interactions in turbulent gas goes beyond the effects of
inducing the drag.}. Schekochihin et al.(2004) and Goldreich \& Sridhar (2006) argued that the plasma viscosity parallel to magnetic 
field can act in the same way as the normal viscosity of unmagnetized fluids (cp Braginskii 1965).  If this is true, the fluids with high magnetic Prandtl number $Pr_m$, i.e. with viscosity much larger than resistivity, are widely spread. In particular, the properties of turbulence in high $Pr_m$ fluid are relevant both to
fully and partially ionized media of the Local Bubble.

\begin{figure*}
\hbox{%
\includegraphics[width=1.9 in]{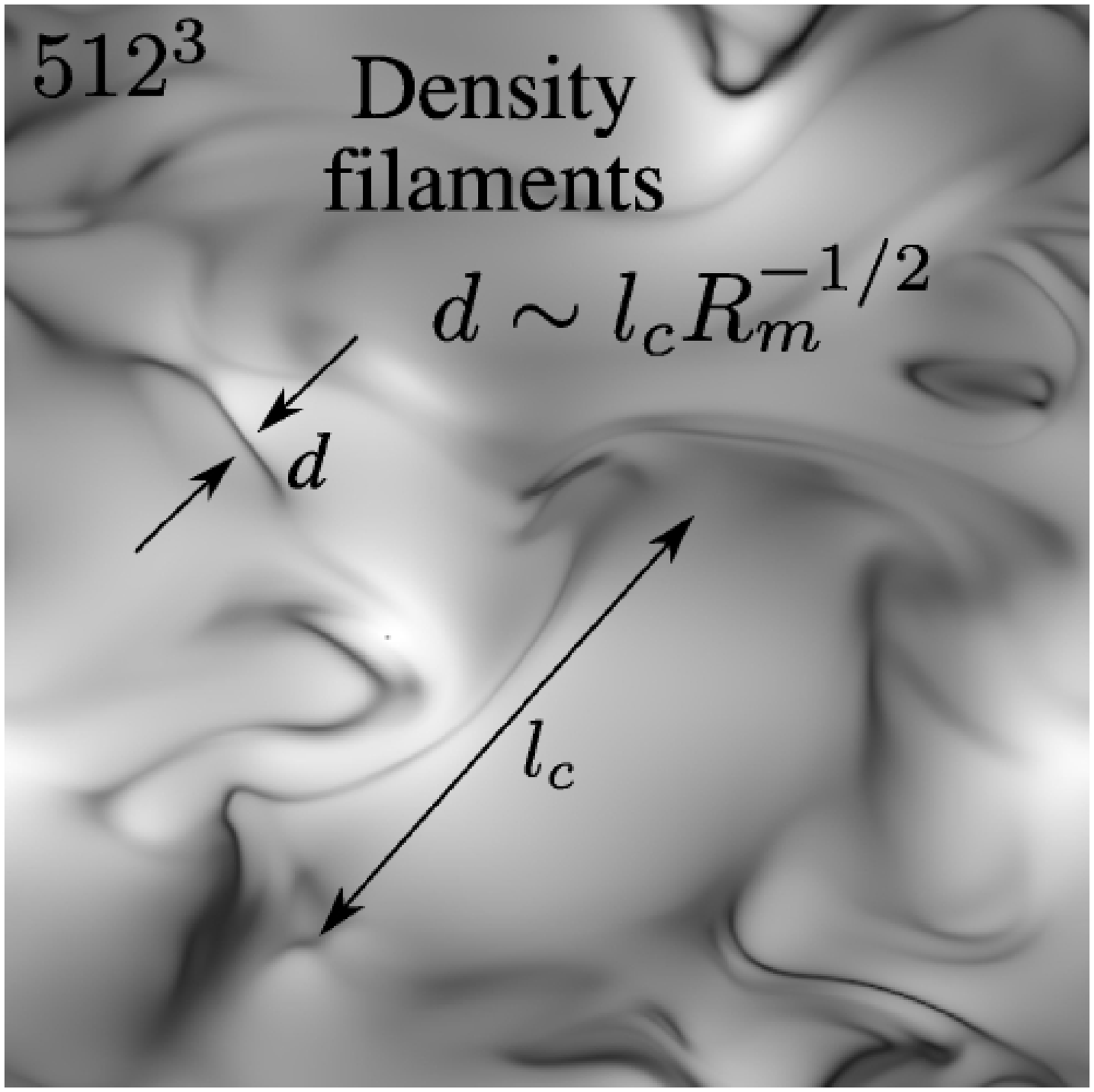}
\hfill
\includegraphics[width=2.1 in]{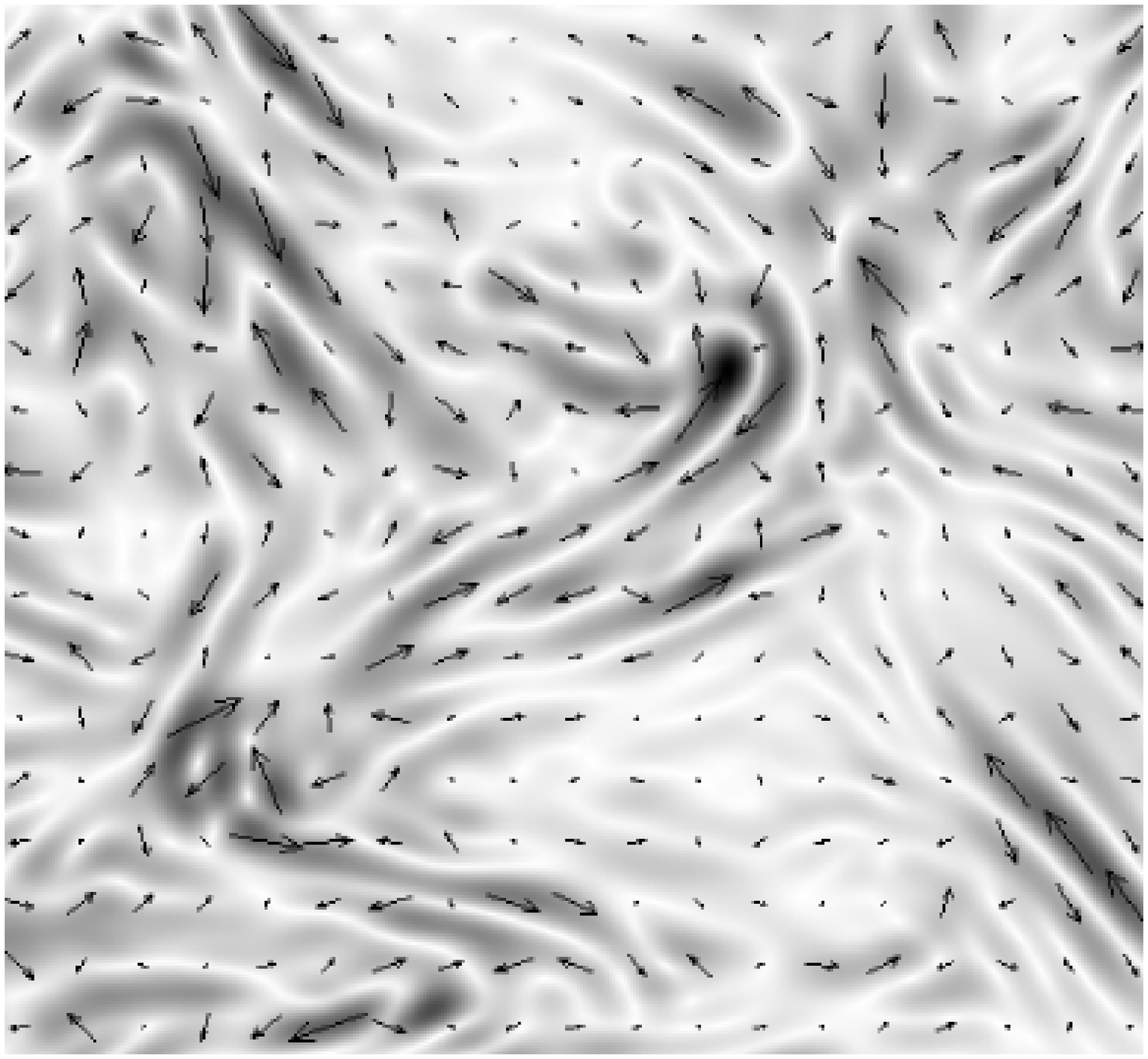}
\hfill
\includegraphics[width=2.5 in]{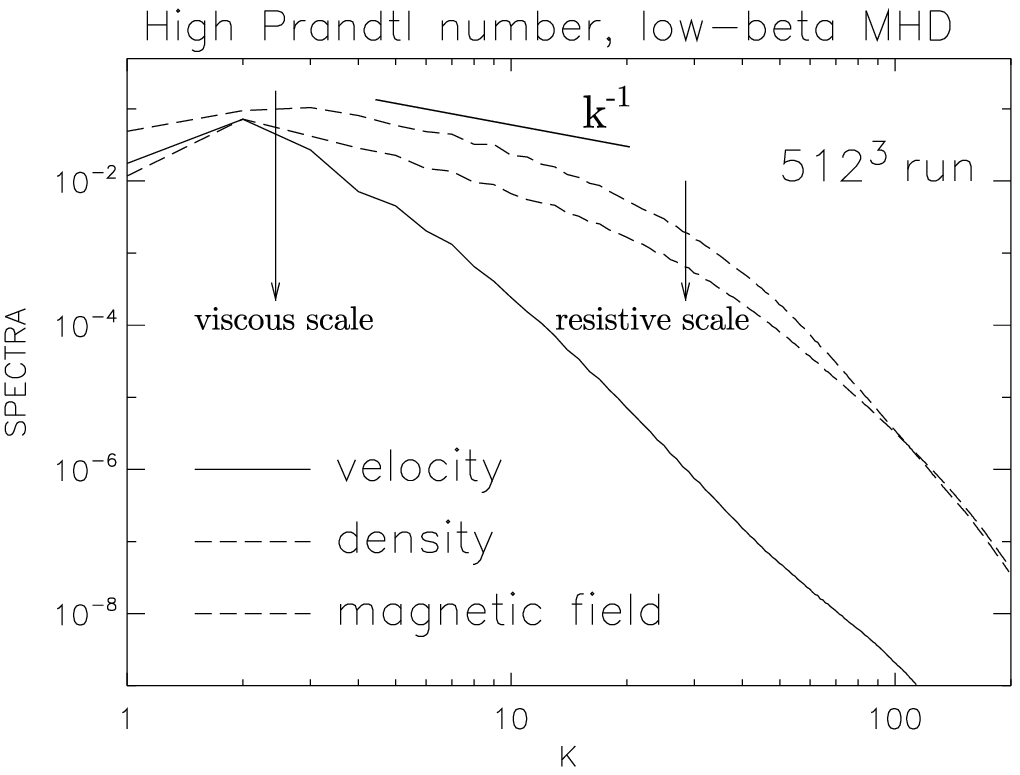}
}
\caption{\small {\it Left}: Filaments of density created by magnetic
compression of the gas in the viscosity-damped regime of MHD turbulence.
Darker regions correspond to higher density. The viscous damping scale $l_c$ is much
larger than the current sheet thickness $d$.  This creates large {\it observed} density contrasts.
{\it Center}: Magnetic reversals (in the plane $\bot$ to mean $\langle{\bf B}\rangle$) that create compressions of density. Darker regions correspond to higher magnetic field.
{\it Right}: Spectra of density and magnetic field are similar,
while velocity is damped. The resistive scale in this regime is not
$L/Rm$ but $L {Rm}^{-1/2}$.  From Beresnyak \& Lazarian. in prep.)}
\label{visc}
\end{figure*}

For high $Pr_m$ fluid Cho, Lazarian \& Vishniac (2002b) reported a new regime of MHD turbulence. 
 Lazarian, Vishniac \& Cho (2004) showed that while the spectrum
of volume-averaged magnetic fluctuations scales as $E_b(k)\sim k^{-1}$
(see Fig.\ref{visc}), the pressure within intermittent magnetic structures
increases with the decrease of the scale $\propto k$ and the
filling factor decreases $\propto k^{-1}$. The magnetic pressure
compresses the gas as demonstrated in Fig.~\ref{visc}. More importantly,
extended current sheets that naturally emerge as magnetic field
fluctuates in the plane perpendicular to the mean magnetic field (see
Fig.~\ref{visc}). It was speculated in Lazarian (2007a) that these current sheets can
account for the origin of the small ionized and neutral structures (SINS) on AU spatial scales 
(Dieter et al. 1976, Heiles 1997, Stanimirovic et al. 2004).

Goldreich \& Sridhar (2006) appealed not to high $Pr_m$ MHD turbulence per se, but
to the generation of the magnetic field in the turbulent plasma (see Schekochihin et al. 2004) to account for the high amplitude, but small scale fluctuations of plasma density observed in the direction of the Galactic center. We believe that the regime of dynamo in Schekochihin et al (2004) and the turbulence in Lazarian et al. (2004) have similarities in terms of the density enhancement that are created. Although in the case of magnetic turbulence with sufficiently strong
mean magnetic field, global reversals, that Goldreich \& Shrindhar (2006) appeal to in compressing plasma, do not happen, the reversals of the magnetic field direction occur in the direction perpendicular to the mean magnetic field. As the mean magnetic field goes to zero, the two regimes get indistinguishable. A systematic study of the density enhancement within the Local Bubble, where more detailed knowledge of
the plasma and magnetic field properties are potentially available, may test the theoretical constructions above. 

\section{Intermittency: small volumes with extreme conditions}

An anisotropic spectrum alone, say $E({\bf k)}\,d{\bf k}$, cannot
characterize MHD turbulence in all its complexity because it involves
only the averaged energy in motions along a particular direction. To
have a full statistical description, one needs to know not only the
averaged spectrum of a physical variable but higher orders as
well. The tendency of fluctuations to become relatively more violent
but increasingly sparse in time and space as the scales get smaller,
so that their influence remains appreciable, is called {\it
intermittency}.   The intermittency increases with the ratio of the size
scales of injection and dissipation of energy, so the very limited
range of scales within numerical simulations may fail to reflect the
actual small scale processes. 

Falgarone et al. (2005,
2006, 2007) and collaborators (Hily-Blant \& Falgarone 2007, Hily-Blant et al. 2007 and refs.  therein) 
attracted the attention of the interstellar community to the potential important implications of
intermittency. A small and transient volume with high temperatures or violent
turbulence can have significant effects on the net rates of processes
within the ISM. For instance, many interstellar chemical reactions
(e.g., the strongly endothermic formation of CH$^+$) might take place
within very intensive intermittent vortices. The aforementioned authors claimed the
existence of the observational evidence for such reactions and heating. If the effects of
intemittency are as strong as Falgarone et al. believe, they should not
be neglected when processes in the Local Bubble are considered.

To get insight into intemittency effects, it is necessary to study high moments of
velocity fluctuations. 
Fortunately, both laboratory and numerical studies demonstrate that
the higher moments of velocity fluctuations can be predicted
remarkably well by the expression derived by She and Leveque (1994)
(see discussion in Lazarian 2006a and ref. therein).  The key
parameter is the spatial dimension of the dissipation structures, $D$,
$\sim1$ for filamentary vortices and $\sim2$ for sheets in MHD (see Muller \& Biskamp 2000,
Cho, Lazarian \& Vishniac 2002,  Boldyrev et
al. 2002). In some instances $D$ can be a fractal dimension in
between. The dimension can be obtained even with low resolution
numerics. This opens an avenue of evaluating the effects of
intermittencies, such as intense local heating or transient effects,
for $Re$ and $Rm$ numbers that will probably never be achieved in
numerical simulations.

The She-Leveque (1994) 
model describes the $p$th power of longitudinal velocity fluctuations
at scale $l$, i.e. $(\delta V_l)^p\sim l^{\xi_p}$, where ${\bf \delta
V_l}\equiv \delta V_l{\bf l}/l$. Here $\delta{\bf V_l}$ is the mean
difference of the local velocity from the average. For
non-intermittent Kolmogorov turbulence, $\xi_p=p/3$ (recall the
well-known result, $\delta V_l\sim l^{1/3}$). In the She-Leveque
(1994) model of turbulence, $\xi_p$ is a more complex function that
depends on (a) the scaling of velocity $\delta V_l\sim l^{\alpha}$,
(b) the energy cascading rate $t_l\sim l^{-\beta}$, and (c) the
dimension of the dissipation structures, $D$. The She-Leveque (1994)
expression is $\xi_p=\alpha\beta
(1-\beta)+(3-D)(1-[1-\beta/(3-D)]^{\alpha\beta})$, with $\alpha
\sim1/3$ and $\beta\sim2/3$ for MHD turbulence.
The principal dependence of $\xi_p$ is on $D$, between 1 and 2.

Dubrulle (1994), She \& Waymire (1995) have figured out that the scaling
exponents $\xi_p$ of the She-Leveque model correspond to so-called
generalized log-Poisson distribution of the local dissipation rate
$\epsilon_l$. With this input, if we constrain
the model parameters  from low resolution numerical
simulations we can calculate the probability
of a fluctuation exceeding a given threshold of the local rate
of deposition of energy, for arbitrary high values of $Re$
and $Rm$. Very importantly, predictions can be made for
the real ISM!

\begin{figure}
\vbox{%
\hskip.2cm\includegraphics[width=5.3cm]{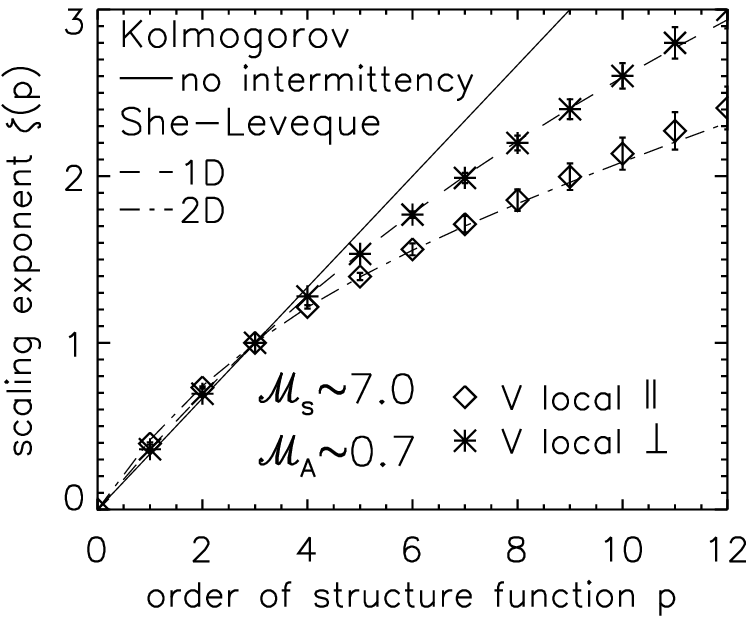}
\vfill
\includegraphics[width=5.5cm]{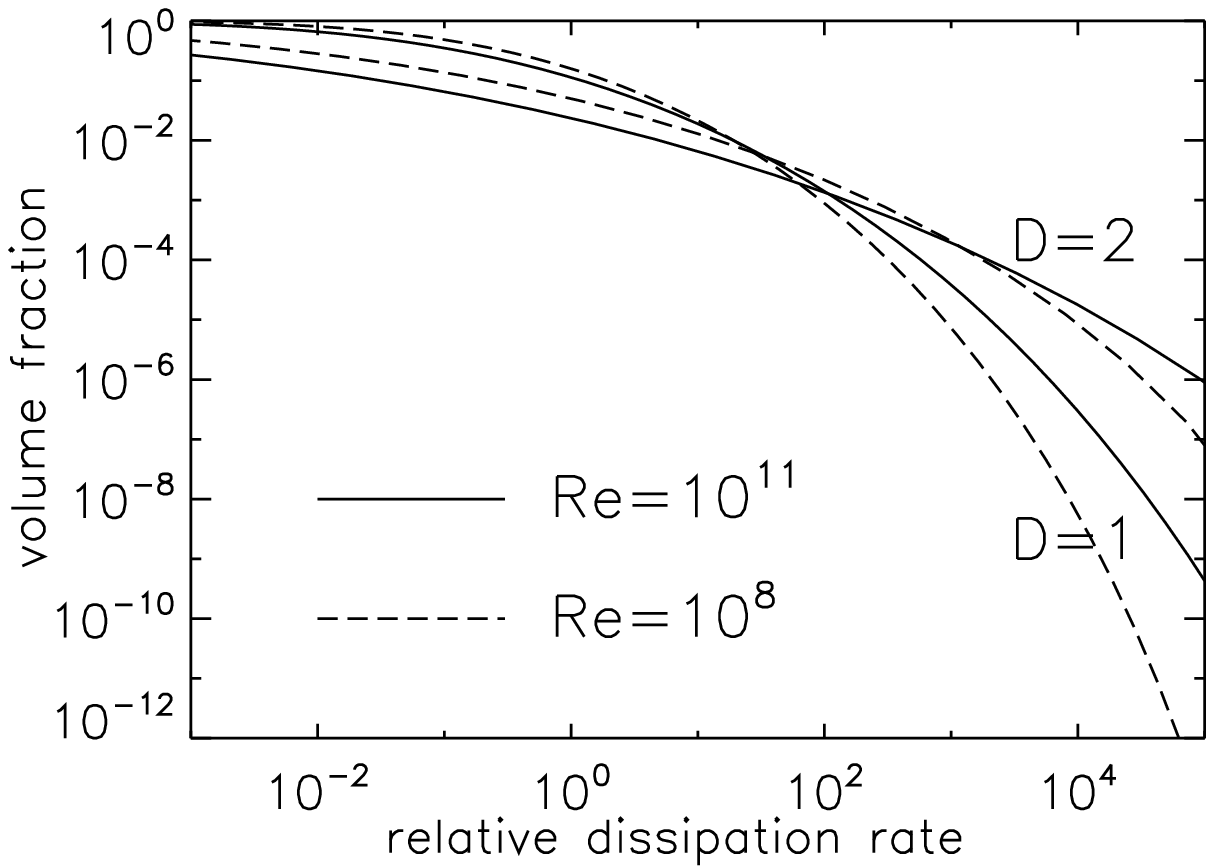}}
\caption{\small {\it Left Panel}: The
intermittencies of velocities in our subAlfvenic, $M_A=0.7$ supersonic
$M_s=7$ MHD simulations. From Kowal \& Lazarian (2008). {\it Right Panel}: Volume fraction with the
dissipation rate is higher than the mean rate for the She-Leveque
model of intermittency with $D=1$ and $2$. From Beresnyak \& Lazarian.}
\label{f1}
\end{figure}

Fig. \ref{f1} shows how to determine $D$. It involves a
simulation (Mach numbers shown) for the component of {\bf V}
perpendicular and also parallel to the local {\bf B}. The dot-dashed
and dashed lines are the respective predictions of the She-Leveque
(1994) expression, with the values of $D$ as indicated. The simulation
clearly provides appropriate values of $D$ for the two components.

The bottom part of Figure \ref{f1} shows calculations in Beresnyak \& Lazarian (2008c)
for the volume fractions of various
dissipation rates (i.e., heating). While the temperatures achieved will depend upon the cooling
functions, some important conclusions are available from the analysis of Figure \ref{f1}.  Indeed,
the model of chemical  reactions by Falgarone et al. requires that a substantial part of the turbulent
cascade energy dissipate in the very intermittent structures. Figure \ref{f1} shows that the bulk of the energy
dissipates within structures where the dissipation rate is higher than the mean value less than the factor
of 100, provided that the She-Leveque model is valid. This provides stringent constraints on what chemistry we
could expect to be induced by intermittent turbulent heating.

Interestingly enough, the case of intermittency studies supports our point of the futility of
the "brute force" numerical approach. For instance, for a
typical ISM injection scale of 50 pc, the Reynolds number can be as high as $Re=10^{11}$.
In comparison, numerical simulations provide $\sim Re^3$
boxes for the present {\it record} resolution of a hydrodynamic simulation
with $4096^3$ boxes.

Even if the intermittency of turbulence is not as important as is in Falgarone et al. model,
its implications for both ISM and Local Bubble plasma may be very important. We see
in Figure \ref{f1} that 10\% of
the energy deposition is localized in just $10^{-2}$\% of the volume!
Such a concentration of energy dissipation in high $Re$ can have many
important consequences on inhomogeneous heating (see Sonnentrucker et
al. 2006). It is clear that further studies in this direction are necessary.

\section{Modification of turbulence by cosmic rays}

It is easy to argue that  the interaction of turbulence and cosmic rays (CRs) is of the most
important processes where properties of MHD turbulence are essential. Indeed, the
interaction of turbulence  with CRs is
a cornerstone of CRs propagation and acceleration models (e.g., Ginzburg 1966, Jokipii
1966, Wentzel 1969, Schlikeiser 2002 and ref. therein).  To account
for the interaction properly, one must know both the scaling of
turbulence and the interactions of turbulence with various waves
produced by CRs.
As we mentioned above, slab Alfvenic modes, which are an essential part of 
many models of cosmic ray propagation (see Goldstein 1976, Bieber et al. 1994,
Matthaeus et al. 2003, Shalchi, Bieber \& Matthaeus 2004, Shalchi 2005, 2006 and references therein), are not observed in
direct MHD simulations. This does not mean that these modes are necessarily 
absent in realistic astrophysical environments. For instance, the influence of the
CRs back onto the turbulence may be important.

The problem is that energetic particles or CRs being an important component in both
the ISM and Local Bubble are not a part of the simulations of MHD turbulence.
 Nevertheless, CRs are
dynamically important and well coupled to the rest of the ISM through
magnetic fields and magnetic turbulence. The spectra of both CRs and
interstellar turbulence show nice power laws, leading R. Jokipii (2001) to suggest a strong interrelation of the two.
It has to be noted that the total pressure of CRs in the ISM is
of the order of the kinetic pressure $\rho v^2$ and, in many cases,
exceed the thermal pressure  by as much as a factor of $10$.
One cannot exclude that CRs may be as important for
ISM simulations as dark matter is for the problems of galaxy formation
and dynamics.

The direct application of the results of MHD turbulence theory to the
problems of the CRs propagation and acceleration is well justified
only when the modification of turbulence by CRs can be neglected. This
has caused controversies since the classical work by Parker
(1965). CRs can induce and are influenced by two instabilities:
($a$) streaming instability (Wentzel 1969, 1974), widely discussed in
the CR literature and frequently referred to as the source of the
postulated scattering ``slab'' , i.e. with ${\bf k \| \bf B}$ Alfvenic
component, and ($b$) the anisotropic pressure kinetic gyroresonance
instability (see Mikhailovskii 1975, Gary 1993, Kulsrud 2004 and refs.
therein), well known to the plasma community.

\begin{figure}
\includegraphics[width=5.4cm]{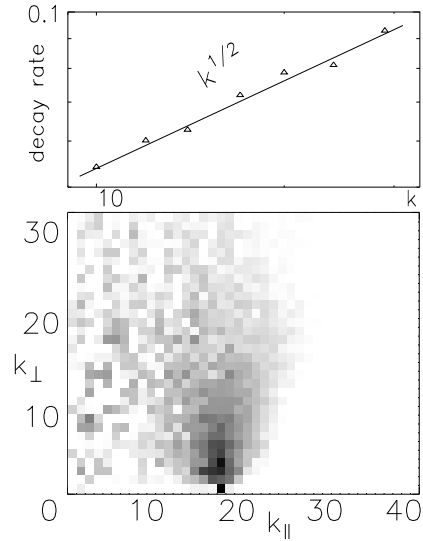}
\caption{\small Decorrelation of a plane, $k_\perp=0$ Alfv\'en wave by
 turbulence. Lower picture shows the energy density of a wave in
 cylindrical k-space. Alfv\'en waves were injected at $k_{\|}=17$.
 From Beresnyak \& Lazarian (2008a).}
\label{SF1}
\end{figure}

The streaming instability, as implied by its name, requires an
anisotropic distribution of particles, while in the interstellar gas,
if we judge from our local measured values, the distribution of CRs is
close to isotropic. In addition, Yan \& Lazarian (2002) noticed that
the streaming instability can be non-linearly suppressed by ambient
turbulence. Farmer \& Goldreich (2004) estimated that galactic
turbulence can efficiently suppress the formation of the ``slab''
Alfvenic waves for protons with relativistic factor
$\gamma>100$. Numerical calculations in Beresnyak \& Lazarian (2008a)
confirmed (see Fig. \ref{SF1}) those predictions.

\begin{figure}
 \includegraphics[width=6.4cm]{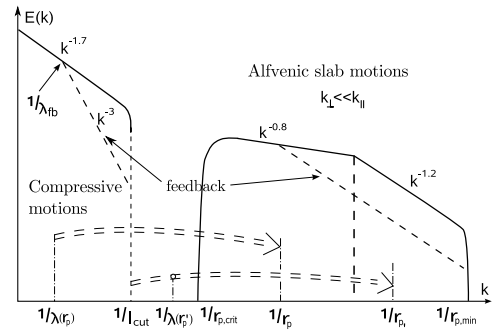}
 \caption{\small Energy density of compressive modes and Alfv\'enic
 slab-type waves, induced by CRs. The energy is transferred from the
 mean free path scale to the CR Larmor radius scale.  See more on
 feedback in Lazarian \& Beresnyak (2006).}
\label{SF2}
\end{figure}

Lazarian \& Beresnyak (2006) research showed that compressible MHD turbulence can generate
kinetic gyroresonance instability that can couple efficiently CRs
with turbulence and backreact on turbulence, modifying its spectrum
(see Fig.~\ref{SF2}). The instability arises from compressing magnetic fields
with gyrating CRs. Indeed, the CR pressure becomes anisotropic as a
consequence of adiabatic invariant conservation, thereby inducing the
instability (Lazarian \& Beresnyak 2006).  The latter gives rise to
``slab'' Alfvenic perturbations that scatter and randomize the CR
momenta via gyroresonance.  As the ``slab'' mode grows, it increases
CR scattering and coupling of CRs and magnetic fields.

Further research of the kinetic gyroresonance instabilities, as well as possibly other instabilities induced by
CRs, should improve our understanding of the importance of the CRs-generated slab Alfven waves. Unlike earlier
models which used somewhat arbitrary (see \S 2) prescriptions for the amplitude of the "slab" Alfvenic
mode (see Bieber et al. 1994), future theories appealing to such modes should be able to evaluate their expected amplitudes (see
first attempts to do this in Lazarian \& Beresnyak 2006). We should also mention, that the slab mode will be subject to damping
while interacting with the surrounding strong Alfvenic turbulence (Yan \& Lazarian 2002, Farmer \& Goldreich 2004, Lazarian \& Beresnyak 2006,
Beresnyak \& Lazarian 2008). Including this damping in Shalchi, Lazarian \& Schlickeiser (2007) improved the fit to the solar wind 
measurement by Ulysses (Gloeckler et al. 1995) and AMPTE spacecraft (Mobius et al. 1998). 

\section{Selected implications of magnetic turbulence} 
The implications of magnetic turbulence for astrophysical fluids, e.g. interstellar medium, are numerous. For instance,
turbulence can heat Diffuse Ionized Gas within the Milky Way (see Minter \& Spangler 1997, Cho et al. 2002), determine the 
evolution of molecular clouds (see McKee \& Ostriker 2007), scatter and accelerate cosmic rays (see Schlickeiser 2002). Below
we consider a couple of {\it selected} examples relevant to the research by the authors to illustrate the progress and problems
on the way of evaluating the effects of turbulence.  

\subsection{Cosmic ray scattering by fast modes of MHD turbulence}
As we mentioned earlier, numerical simulations of MHD turbulence supported the GS95 model of turbulence,
which does not have the "slab" Alfvenic modes that produced most of the scattering in the earlier models
of CR propagation. Can the turbulence that does not appeal to CRs back-reaction (see \S 4) produce 
efficient scattering?

\begin{figure*} [h!t] 
{\centering \leavevmode
\includegraphics[width=2.5in]{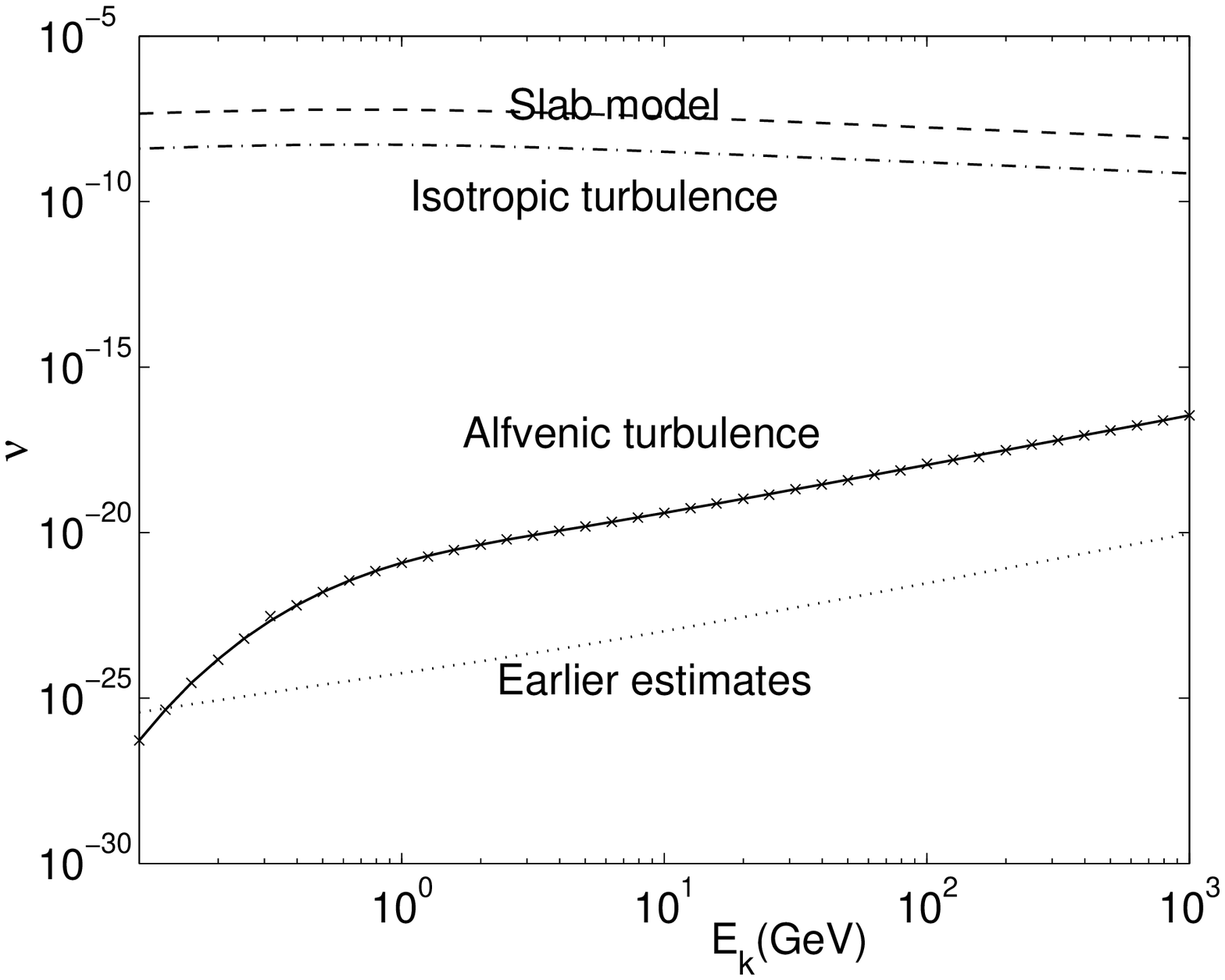} 
\hfil 
\includegraphics[width=2.5in]{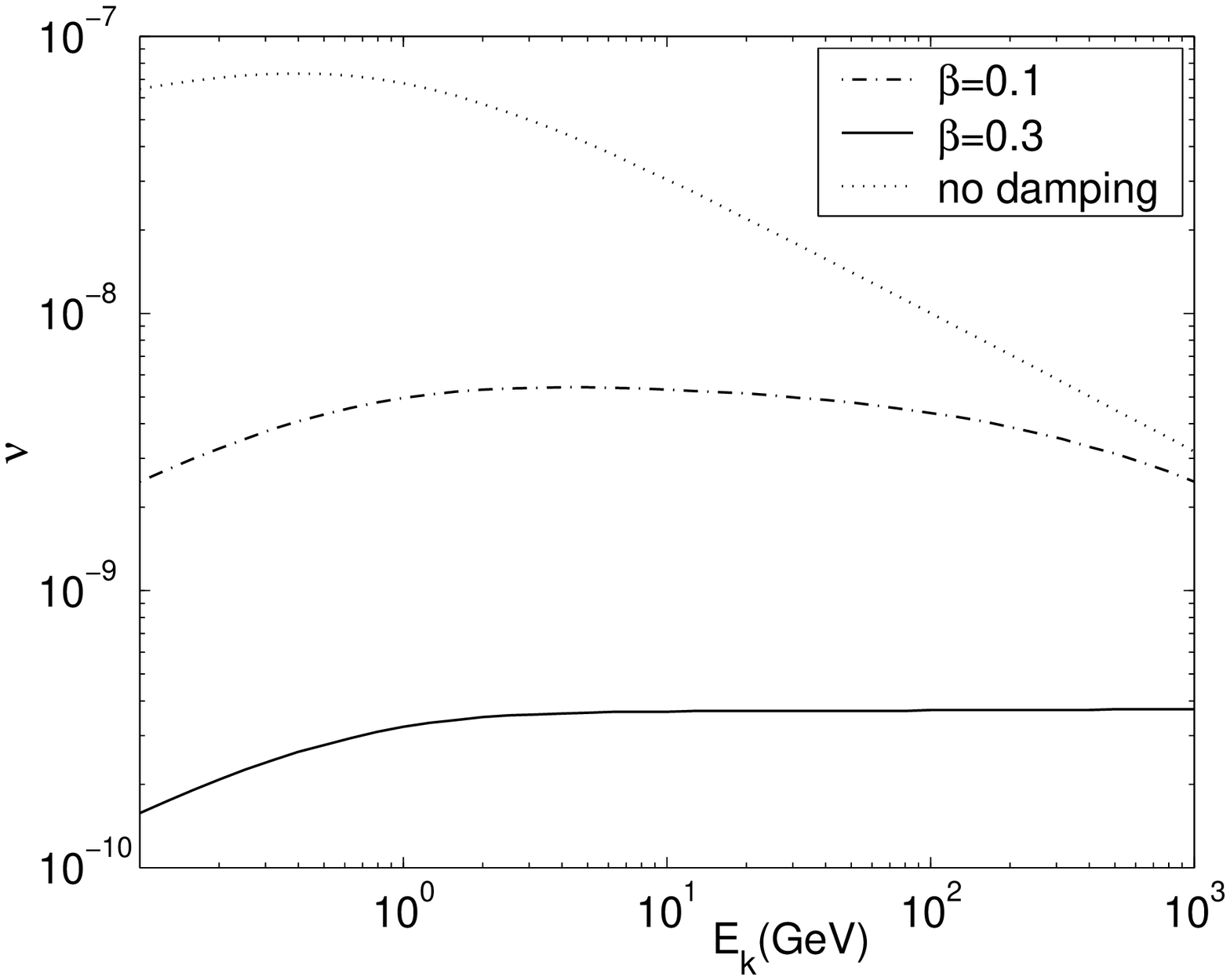}
}
{\centering \leavevmode
\includegraphics[width=2.5in]{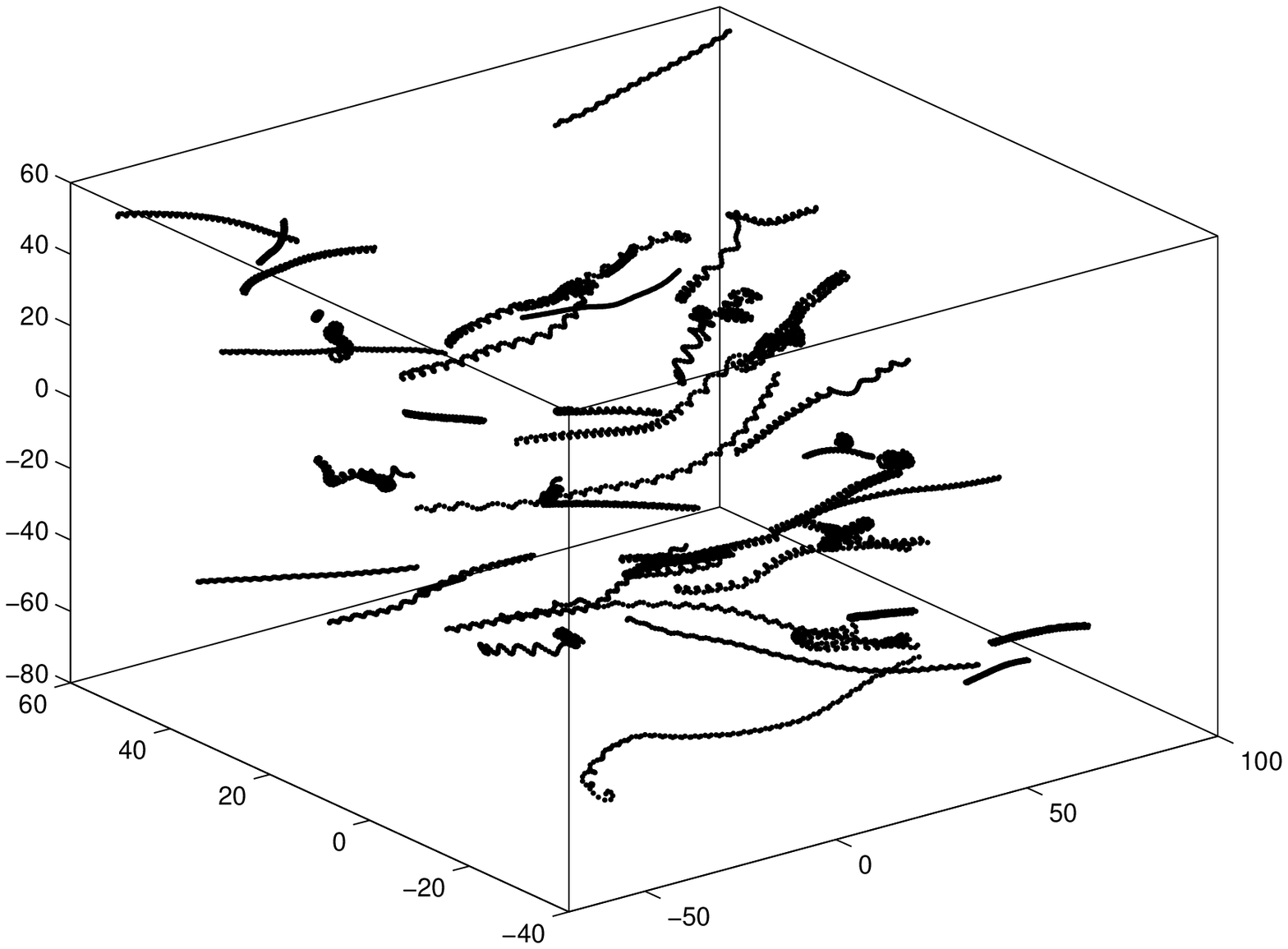} 
\hfil 
\includegraphics[width=2.5in]{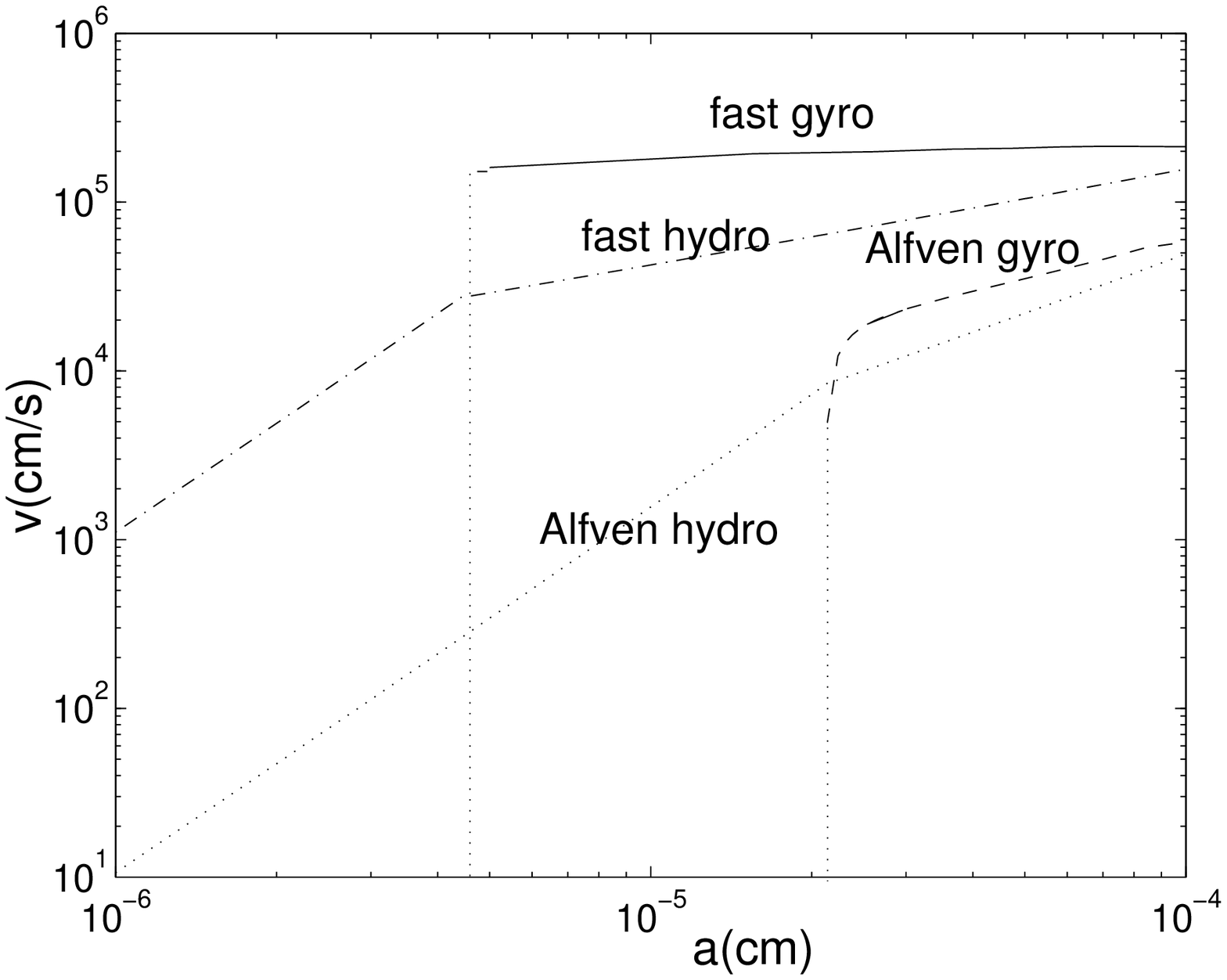}
} 
\caption{ {\it Implications of Interstellar Turbulence for the Cosmic Rays
and Interstellar Dust.}  
{\it Upper left:} Rate of CR scattering by
Alfven waves versus CR energy.  The lines at the top of the figure are
the accepted estimates obtained for Kolmogorov turbulence. The dotted
curve is from Chandran (2000). The analytical calculations are given
by the solid line with our numerical calculations given by
crosses. {\it Upper right:} The rate of CR scattering ($\nu$) by fast
modes in magnetically dominated plasma. The rate of scattering depends
on damping of the fast waves (see , which in turn depends on the ratio of
gaseous to magnetic pressure ($\beta= P_{gas}/P_{mag}$). {\it Lower
left:} Individual trajectories of CRs tracked by the  Monte Carlo scattering
code. {\bf B} is obtained through 3-D simulations of MHD
turbulence. These calculations provide estimates of CR
diffusion.  {\it Lower right:} Velocities of charged dust grains in
cold neutral media (CNM).  Gyroresonance with fast modes (``fast gyro'')
dominates for large grains, while hydrodynamic drag (``fast hydro'')
for small grains. The cutoffs for Alfven and
fast gyro (vertical lines) are due to MHD 
turbulence damping caused by neutral-ion collisions. From Yan \& Lazarian 2002, 2003, 2008.}
\label{impl}
\end{figure*}

In the models of ISM turbulence (Armstrong et al. 1994,  McKee \& Ostriker 2007), where the injection happens at large scale, 
fast modes were identified as a scattering agent for cosmic rays in interstellar medium
  Yan \& Lazarian (2002, 2004).
These works made use of the quantitative description of turbulence
obtained in CL02  to calculate
the scattering rate of cosmic rays. The results are shown in
Fig.~\ref{impl}. For instance, the scattering rate of relativistic protons by
Alfvenic turbulence was shown to be nearly $10^{10}$ times lower than
the generally accepted estimates obtained assuming the Kolmogorov
scaling of turbulence.  Although this estimate is $10^{4}$ times
larger than that obtained by Chandran (2000), who employed
GS95 ideas of anisotropy, but lacked the quantitative
description of the eddies, it is clear that for most interstellar
circumstances the Alfvenic scattering is suppressed.  The low efficiency of scattering by
Alfvenic modes arise from high anisotropy of the modes at the scales of cosmic ray gyroradius.

YL02 showed that
the scattering by fast modes, which are isotropic (CL02), dominates
(see Fig.~\ref{impl}). However, fast modes are subject to both
collisional and collisionless damping\footnote{On the basis of weak turbulence theory, Chandran (2005) has argued that high-frequency 
fast waves, which move mostly parallel to magnetic field, generate Alfven waves also moving mostly parallel to magnetic field. We expect
that the scattering by thus generated Alfven modes to be similar to the scattering by the fast modes created by them. Therefore
we expect that the simplified approach adopted in Yan \& Lazarian (2004) and the papers that followed to hold.}, which was taken into account in Yan \& Lazarian (2004).
More recent studies of cosmic ray propagation and acceleration that explicitly appeal to the effect of
the fast modes include Cassano \& Brunetti 2005, Brunetti 2006, Brunetti \& Lazarian (2007), Yan \& Lazarian (2008) and Yan, Lazarian \& Petrosian (2008).
Incidentally, fast modes have been also identified as primary agents for the acceleration of charged dust particles (Yan \& Lazarian 2003,
Yan, Lazarian \& Draine 2004).

\subsection{Turbulent reconnection and cosmic ray acceleration}
Magnetic reconnection can be associated with the ability of magnetic flux tubes to change their topology, while
being submerged within conducting fluids (see Biskamp 1996). Due to high numerical diffusivity of present-day simulations, reconnection
is fast there, which, for instance, means that magnetic fields in the ISM, Local Bubble and solar wind simulations
change their topology fast. Is this true for the real astrophysical circumstances?

Recent years have been marked by a substantial progress in simulations of collisionless reconnection (see Shay \& Drake 1998,
Bhattacharjee et al. 2003, Drake et al. 2006 and references therein). However, while
the researchers argue whether Hall MHD or fully kinetic description (Daughton 2006)  is necessary, one statement is definitely true. If magnetic reconnection
is only fast in collisionless environments, most of the MHD simulations, e.g. of interstellar medium, accretion disks, stars, where the
environment is collisional, are in error. We shall argue below that this radical conclusion may not be true and the reconnection is
also fast in most astrophysical collisional environments.  

\begin{figure}
\includegraphics[width=.5\textwidth]{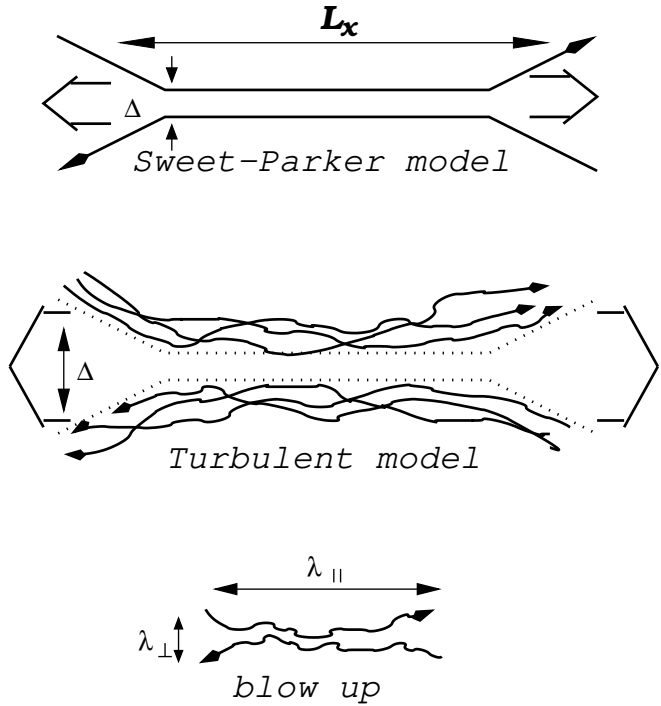}
\caption{{\it Upper plot}: 
Sweet-Parker model of reconnection. The outflow
is limited by a thin slot $\Delta$, which is determined by Ohmic 
diffusivity. The other scale is an astrophysical scale $L\gg \Delta$.
{\it Middle plot}: Turbulent reconnection model that accounts for the 
stochasticity
of magnetic field lines. The outflow is limited by the diffusion of
magnetic field lines, which depends on field line stochasticity.
{\it Low plot}: An individual small scale reconnection region. The
reconnection over small patches of magnetic field determines the local
reconnection rate. The global reconnection rate is substantially larger
as many independent patches come together.}
\label{recon}
\end{figure}

Lazarian \& Vishniac (1999, henceforth LV99) considered  turbulence as the agent that makes magnetic reconnection fast.
The scheme proposed in LV99 there differs appreciably from the earlier attempts to enhance reconnection via turbulence (Speiser 1970,
Jacobson \& Moses 1984,  Matthaeus \& Lamkin 1985, Bhattacharjee \& Hameiri 1986, Hameiri \& Bhattacharjee 1987, Straus 1988, see 
Lazarian et al. 2004 for a detailed comparison).
 The scheme proposed is a generalization of the   
Sweet-Parker scheme (see Fig.~\ref{recon}). The
problem of the Sweet-Parker model is that the reconnection is negligibly slow for any realistic 
astrophysical conditions. However, astrophysical magnetic fields are generically turbulent.

LV99 consider the case in which there exists a large scale,
well-ordered magnetic field, of the kind that is normally used as
a starting point for discussions of reconnection.  
In addition, we expect that the field has some small scale `wandering' of
the field lines.  On any given scale the typical angle by which field
lines differ from their neighbors is $\phi\ll1$, and this angle persists
for a distance along the field lines $\lambda_{\|}$ with
a correlation distance $\lambda_{\perp}$ across field lines (see Fig.~\ref{recon}).

The modification of the global constraint induced by mass conservation
 in the presence of
a stochastic magnetic field component 
is self-evident. Instead of being squeezed from a layer whose
width is determined by Ohmic diffusion, the plasma may diffuse 
through a much broader layer, $L_y\sim \langle y^2\rangle^{1/2}$ (see Fig.~\ref{recon}),
determined by the diffusion of magnetic field lines.  This suggests
an upper limit on the reconnection speed of 
$\sim V_A (\langle y^2\rangle^{1/2}/L_x)$. 
This will be the actual speed of reconnection if
the progress of reconnection in the current sheet does not
impose a smaller limit. The value of
$\langle y^2\rangle^{1/2}$ can be determined once a particular model
of turbulence is adopted, but it is obvious from the very beginning
that this value is determined by field wandering rather than Ohmic
diffusion as in the Sweet-Parker case.

What about limits on the speed of reconnection that arise from
considering the structure of the current sheet?
In the presence of a stochastic field component, magnetic reconnection
dissipates field lines not over their  entire length $\sim L_x$ but only over
a scale $\lambda_{\|}\ll L_x$ (see Fig.~\ref{recon}), which
is the scale over which magnetic field line deviates from its original
direction by the thickness of the Ohmic diffusion layer $\lambda_{\perp}^{-1}
\approx \eta/V_{rec, local}$. If the angle $\phi$ of field deviation
does not depend on the scale, the local
reconnection velocity would be $\sim V_A \phi$ and would not depend
on resistivity. In LV99 it is taken into account that $\phi$ does depend on scale (see \S 2).
Therefore the {\it local} 
reconnection rate $V_{rec, local}$ is given by the usual Sweet-Parker formula
but with $\lambda_{\|}$ instead of $L_x$, i.e. $V_{rec, local}\approx V_A 
(V_A\lambda_{\|}/\eta)^{-1/2}$.
It is obvious from Fig.~\ref{recon} that $\sim L_x/\lambda_{\|}$ magnetic field 
lines will undergo reconnection simultaneously (compared to a one by one
line reconnection process for
the Sweet-Parker scheme). Therefore the overall reconnection rate
may be as large as
$V_{rec, global}\approx V_A (L_x/\lambda_{\|})(V_A\lambda_{\|}/\eta)^{-1/2}$.
Whether or not this limit is important depends on
the value of $\lambda_{\|}$.  

\begin{figure*}
\includegraphics[width=.3\textwidth]{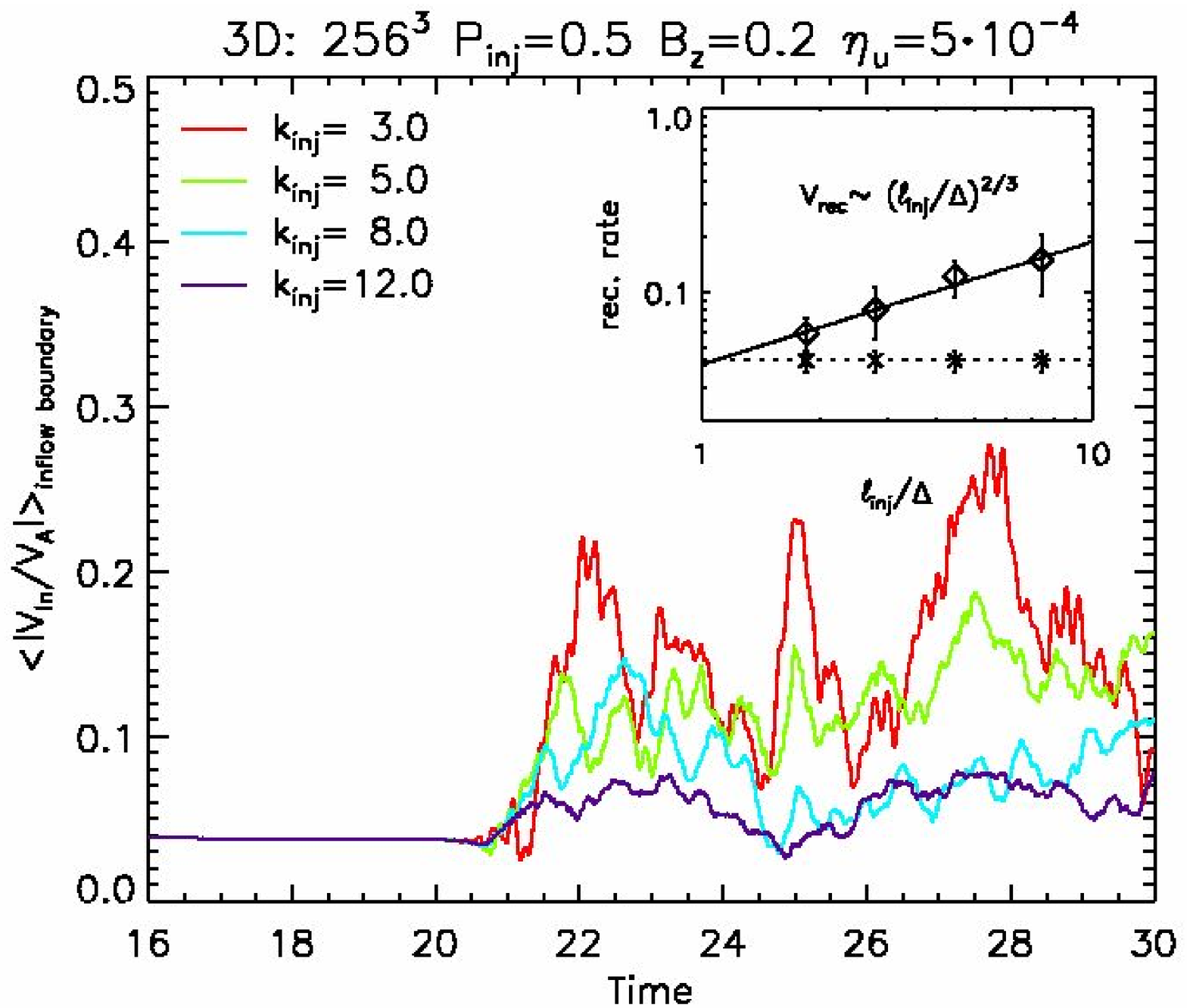}
\includegraphics[width=.3\textwidth]{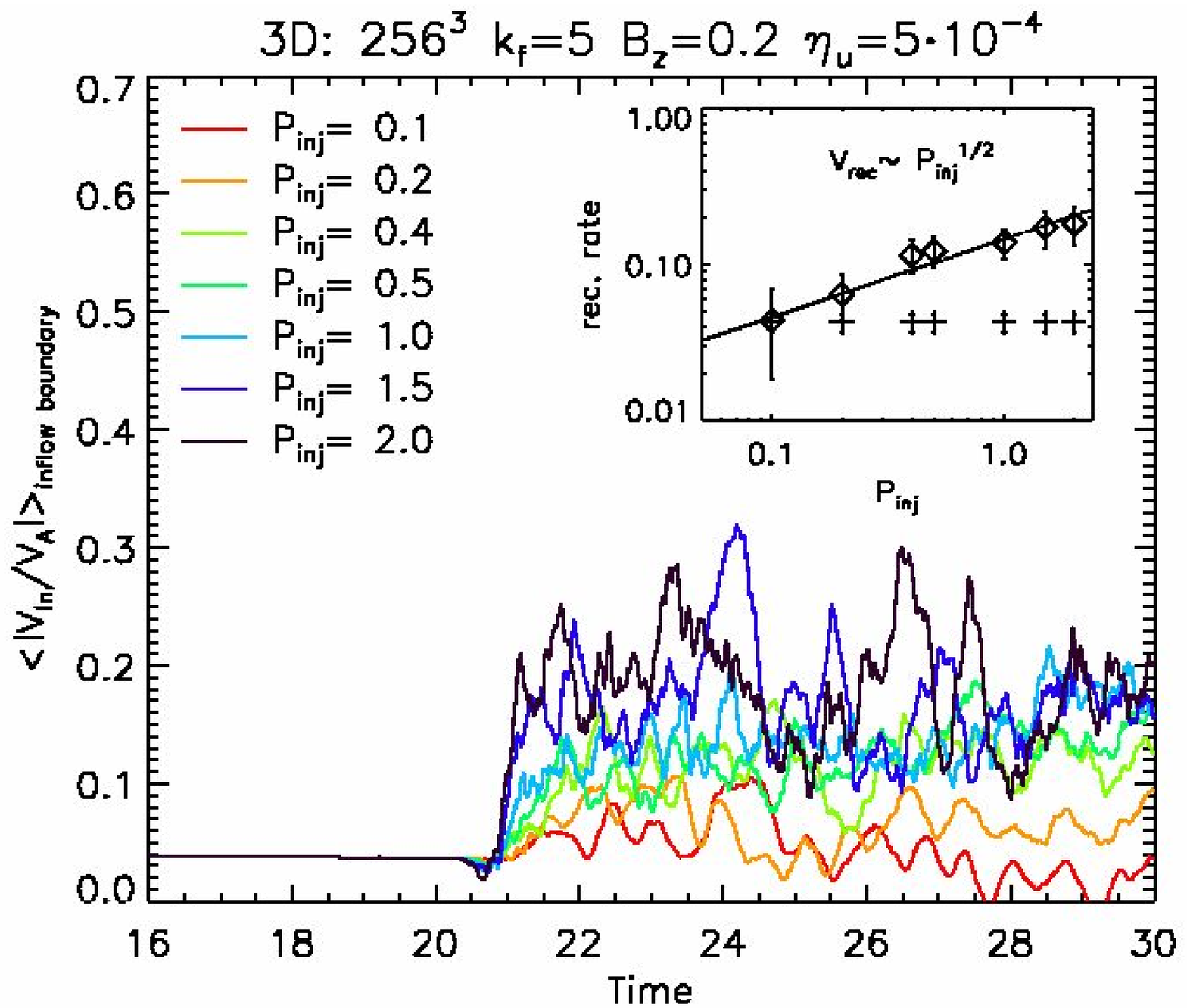}
\includegraphics[width=0.3\textwidth]{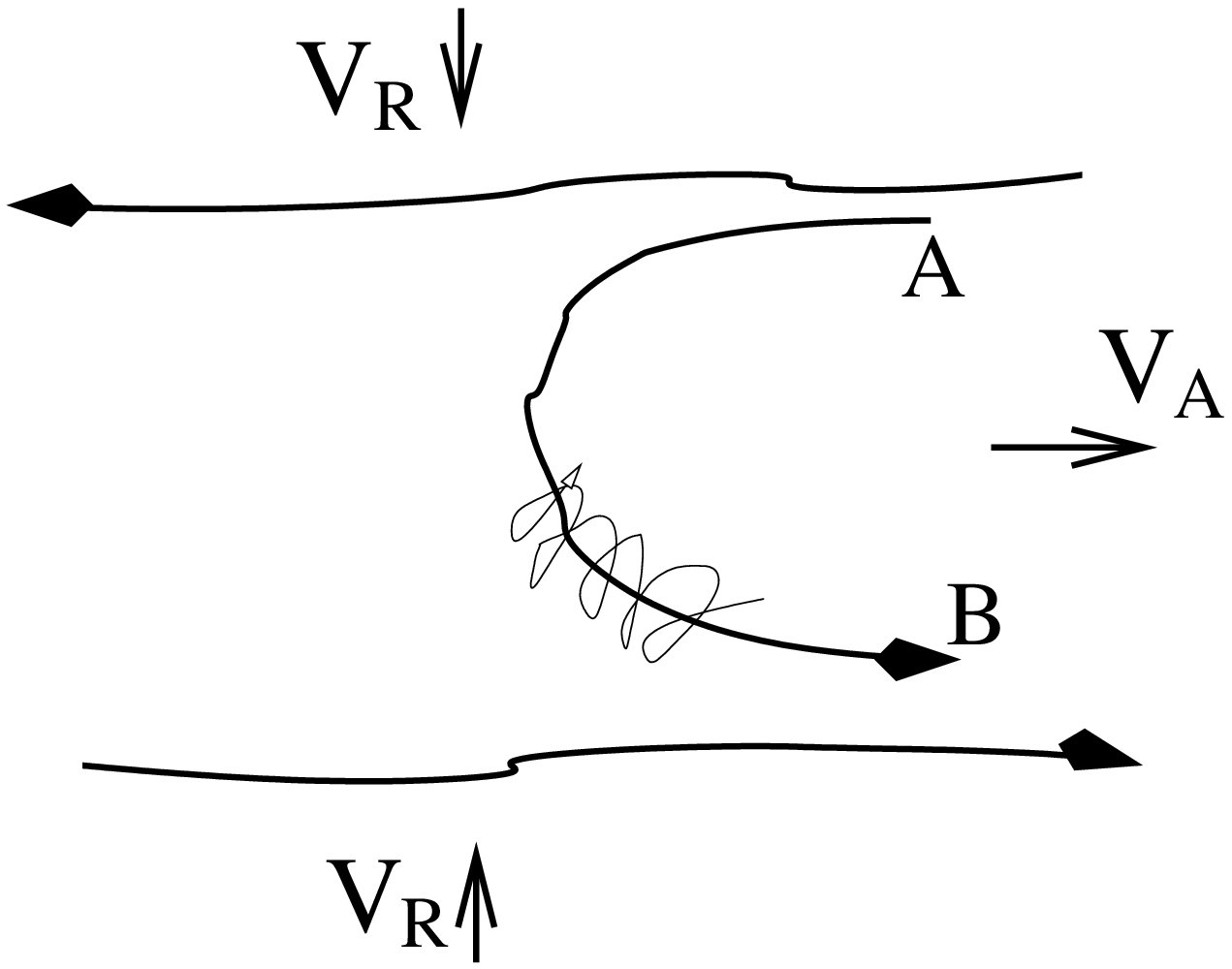}
\caption{ {\it Left panel} Testing of the model of 3D turbulent reconnection with MHD simulations in Kowal et al. (2008)
The results, consistently with the LV99 model, show an increase of the reconnection rate with the increase
of the injection scale of the turbulence. The reconnection is bursty, as is expected for the turbulence model. The reconnection rate is normalized to the Alfvenic velocity, time is given in the units of Alfvenic crossing times. Initially the turbulence is slow, but its velocity increases as the turbulence is injected. {\it Central Panel}. The same as for the left panel, but the injection energy is being changed.
The results are also consistent with the LV99 model. {\it Right panel}: 
Cosmic rays spiral about a reconnected magnetic
field line and bounce back at points A and B. The reconnected
regions move towards each other with the reconnection velocity
$V_R$. The advection of cosmic rays entrained on magnetic field
lines happens at the outflow velocity, which is in most cases
of the order of $V_A$. Bouncing at points A and B happens
because either of streaming instability or turbulence in the
reconnection region.}
\label{recon2}
\end{figure*}

The relevant values of $\lambda_{\|}$ and $\langle y^2\rangle^{1/2}$ 
depend on the magnetic field statistics. This
calculation was performed in LV99 using the GS95 model
of MHD turbulence providing 
the upper limit on the reconnection speed:
\begin{equation}
V_{r, up}=V_A \min\left[\left({L_x\over l}\right)^{\frac{1}{2}}
\left({l\over L_x}\right)^{\frac{1}{2}}\right]
\left({v_l\over V_A}\right)^{2},
\label{main}
\end{equation}
where $l$ and $v_l$ are the energy injection scale and
turbulent velocity at this scale respectively.
In LV99 other processes that can impede
reconnection were found to be less restrictive. For
instance, the tangle of reconnection field lines crossing the
current sheet will need to reconnect repeatedly before individual
flux elements can leave the current sheet behind.  The rate at which
this occurs can be estimated by assuming that it constitutes the
real bottleneck in reconnection events, and then analyzing each
flux element reconnection as part of a self-similar system of
such events.  This turns out to impede the reconnection.  
As the result, LV99 
concludes that (\ref{main}) is not only an
upper limit, but is the best estimate of the speed of reconnection. The model
has been recently tested numerically in Kowal et al. (2008) (see Fig. \ref{recon2}). 
The thick current sheets observed during the  2003 November 4 Coronal Mass Ejection reported
in Ciaravella \& Raymond (2008) are also consistent with the LV99 model.

The most interesting process is the first-order Fermi acceleration
that is intrinsic to the turbulent reconnection. To understand it
consider a particle entrained on a reconnected 
magnetic field line (see Figure\ref{recon}). This particle
may bounce back and forth between magnetic mirrors formed by oppositely
directed magnetic fluxes moving towards each other with the velocity
$V_R$. Each of such bouncing will increase the energy of a particle
in a way consistent with the requirements of the first-order Fermi
process. The interesting property of this mechanism that potentially
can be used  to test the idea observationally is that the 
resulting spectrum is different from those arising from shocks.
Gouveia Dal Pino \& Lazarian (2003) used this mechanism of particle acceleration\footnote{The mechanism
has physical similarities to the acceleration mechanism that was proposed later for electrons by Drake et al. (2006b).
In Drake's mechanism, similarly, to the Matthaeus, Ambrosiano \& Goldstein (1984) mechanism, however, the process of acceleration happens within 2D contracting loops. 
For LV99 model of reconnection the generic configuration of magnetic field are contracting  spirals.}  to explain
the synchrotron power-law spectrum arising from the flares of the
microquasar GRS 1915+105. Note, that the mechanism acts in the
Sweet-Parker scheme as well as in the scheme of turbulent reconnection.
However, in the former the rates of reconnection and therefore the
efficiency of acceleration are marginal in most cases.

\subsection{Perpendicular diffusion and subdiffusion}

Modeling of clouds submerged in hot plasma present one of the challenges to the research of the Local Bubble physics.
It is easy to see that the diffusion heat perpendicular to magnetic field is very slow if the field is laminar. Incidentally, the
same is true in terms of CR diffusion, while the observations indicate that the diffusion of the CRs perpendicular to magnetic
field in the Galaxy is reduced just by a factor $\sim 3$ compared to the parallel diffusion (Jokipii 1999). How could this be?

The reader may already guess that, similar to the reconnection problem discussed above, magnetic field wandering
may allow the particles to diffuse (Jokipii \& Parker 1969,  Giacalone, J., \& Jokipii, J.~R.\ 1999). Analogously with the reconnection example, one may
expect that the expression for the field wandering would depend on the model of turbulence accepted. Within the GS95 model
of turbulence and assuming that the injection turbulent velocity is equal to the Alfven one, the calculations for heat diffusion were performed in Narayan \& Medvedev 2001 A general case of arbitrary injection velocity was considered in Lazarian (2006). The heat diffusion in plasma was found to be a function of the Alfven Mach number $M_A$ defined as the ratio of the turbulent velocity at the injection scale to the Alfven speed in the medium. For the case of strong mean magnetic field, i.e. for $M_A<1$, and mean free path of a electron $\lambda$ less than the injection scale of the turbulence $L$, the heat diffusion coefficient was obtained to be $1/3 M_A^4 \kappa_{Spitzer}$, where $\kappa_{Spitzer}=\lambda v_{therm}$ is the usual Spitzer heat diffusion coefficient of unmagnetized plasma. The factor $1/3$ there reflects the 1D nature of diffusion along
magnetic field lines, while the $M_A^4$ power\footnote{There is a wrong power in the original Lazarian (2006) paper, which was  corrected
in Lazarian (2007).} reflects the inefficiency of magnetic field line wandering in the presence of the strong mean magnetic
field. Similarly, for superAlfvenic turbulence, i.e. for $M_A>1$ one gets the heat diffusion coefficient $1/3 \kappa_{Spitzer}$ if the mean free path
of an electron $\lambda < LM_A^{-3}$ and gets $1/3 (L M_A^{-3}/\lambda) \kappa_{Spitzer}$ in the opposite regime.

\begin{figure}
 \includegraphics[width=.5\textwidth]{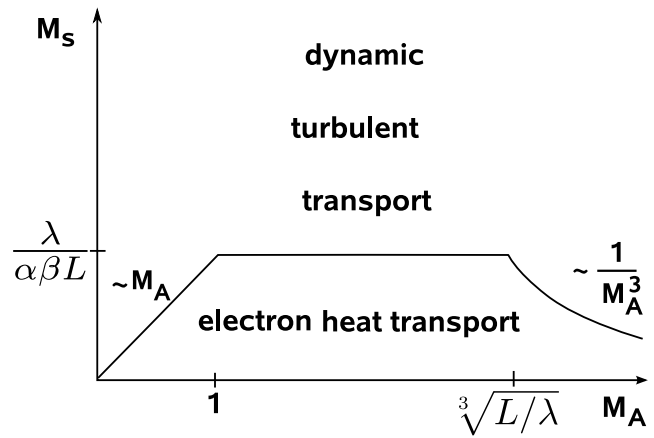}
\caption{ The parameter spaces for the advection of heat by turbulent motions and heat conduction by plasma electrons. Sonic Mach number $M_s$ is plotted against the Alfven Mach number $M_A$. The mean free path of electrons is $\lambda$, while $\alpha$ is the square root of the
electron to proton mass ratio, i.e. $(m_e/m_p)^{1/2}$, $\beta$ is a numerical factor $\approx 4$. From Lazarian 2006, 2007.}
\label{diff}
\end{figure}

Cho et al. (2003) noticed that in magnetized turbulence, the turbulence eddies should induce in plasma the heat advection with the effective diffusion coefficient $2/3 LV_L$, where the factor of $1/3$ is similar to the advective diffusion in hydrodynamics (Lesieur 1990) and the factor of
$2$ takes into account that both electrons and protons participate in the process. The relative importance of the two diffusion processes is
exemplified in Figure \ref{diff}, where the parameter space for the dominance of the heat advection by turbulent motions and the dominance of
the heat transfer by electrons are defined. These results for the fully ionized plasma can be used for parameterizing heat transport in the astrophysical codes. They also can easily be generalized for the partially ionized gas. Note, that
following individual particles in simulations may be prohibitively expensive (see Tilley \& Balsara 2006). Moreover, the actual structure of magnetic field is distorted by the limited numerical resolution.

The last point of magnetic field structure in numerical simulations is also important for the CRs diffusion perpendicular to magnetic field.
Take, for instance, the issue of subdiffusion, which is the non-diffusive behavior arising from the CRs retracing their trajectories, as a result
of the backward scattering (see Kota \& Jokipii 2000, Qin et al. 2002, Webb et al. 2006). In the case of the unrealistic turbulence with only one scale of turbulent motions $l_{turb}$, the retracing 
of particles stops as soon as the particles diffuse the so-called Rechester \& Rosenbluth (1978) length, which is $l_{turb} \ln(l_{turb}/r_{Lar})$,
where $r_{Lar}$ is the Larmor radius of a charged particle (see also Chandran \& Cowley 1998). In the case of realistic turbulence with a range of scales one should use the dissipation scale instead of the $t_{turb}$ (Narayan \& Medvedev 2001, Lazarian 2006, Yan \& Lazarian 2008). The
corresponding scale for the Alfvenic turbulence in the ISM may be less than $l_{crit}\approx 10^6$ km, thus one should expect subdiffusion only at scales less than this. For the solar wind turbulence  $l_{crit}$ is even smaller, of the order of $10^3$ km. On the contrary, for the numerical simulations with a limited inertial range, the subdiffusion may be much more important (see Qin et al. 2002).  

\section{Studying magnetic turbulence in Interplanetary medium using comets}

As we illustrated with selected examples above, the modeling of turbulent astrophysical fluids is far from being simple. Therefore, observational
testing is essential. We are fortunate to have in situ measurements of magnetic field in the interplanetary medium (see Opher et al. 2007). The advantage of direct studies of magnetic perturbations by spacecrafts has been explored through many important missions.  Such studies, unlike numerical ones,
may deliver information about the actual magnetic turbulence at high $Re$ and $Rm$ numbers. However, the spacecraft measurement are rather expensive. Are there any other cost-effective ways to study magnetic turbulence in interplanetary medium?

Below we present a new way to explore the turbulence in interplanetary medium by using the alignment
of sodium atoms ejected from comets. Atomic alignment of atoms in their ground state was an effect studied in the middle of
the previous century in laboratories, in relation to maser research (Hawkins 1955, Kastler 1957).  The alignment of atoms is understood in terms of their non-equilibrium distribution of ground level substates. Therefore, to be aligned in the ground state the atom in question should
have either fine or hyper-fine structure. The non-equilibrium distribution arises from radiative pumping, while the magnetic field realigns 
atoms inducing their Larmor precession. As the substates of the ground level are long-lived, even a small magnetic field modifies the distribution
and therefore the polarization arising from the aligned atoms (see Yan \& Lazarian 2007 for more details). 

Comets are known to have Sodium tails and Sodium is an atom that can be aligned by radiation and realigned by solar wind magnetic fields. This opens an opportunity of studying magnetic fields in the solar wind from the ground, but tracing the polarization of the Sodium line.
At the moment this is a suggestion supported by the synthetic ground based observations. For our synthetic observations we used a model of magnetic field in Liu et al. (2008). It employed the space weather model developed by University of Michigan, namely,
The Space Weather Modeling Framework (SWMG) (Toth et al 2005). More specifically, this is a solar corona and inner heliosphere
model that extends the description of media from the solar surface to 1AU. 

\begin{figure*}
{
\includegraphics[width=.3\textwidth]{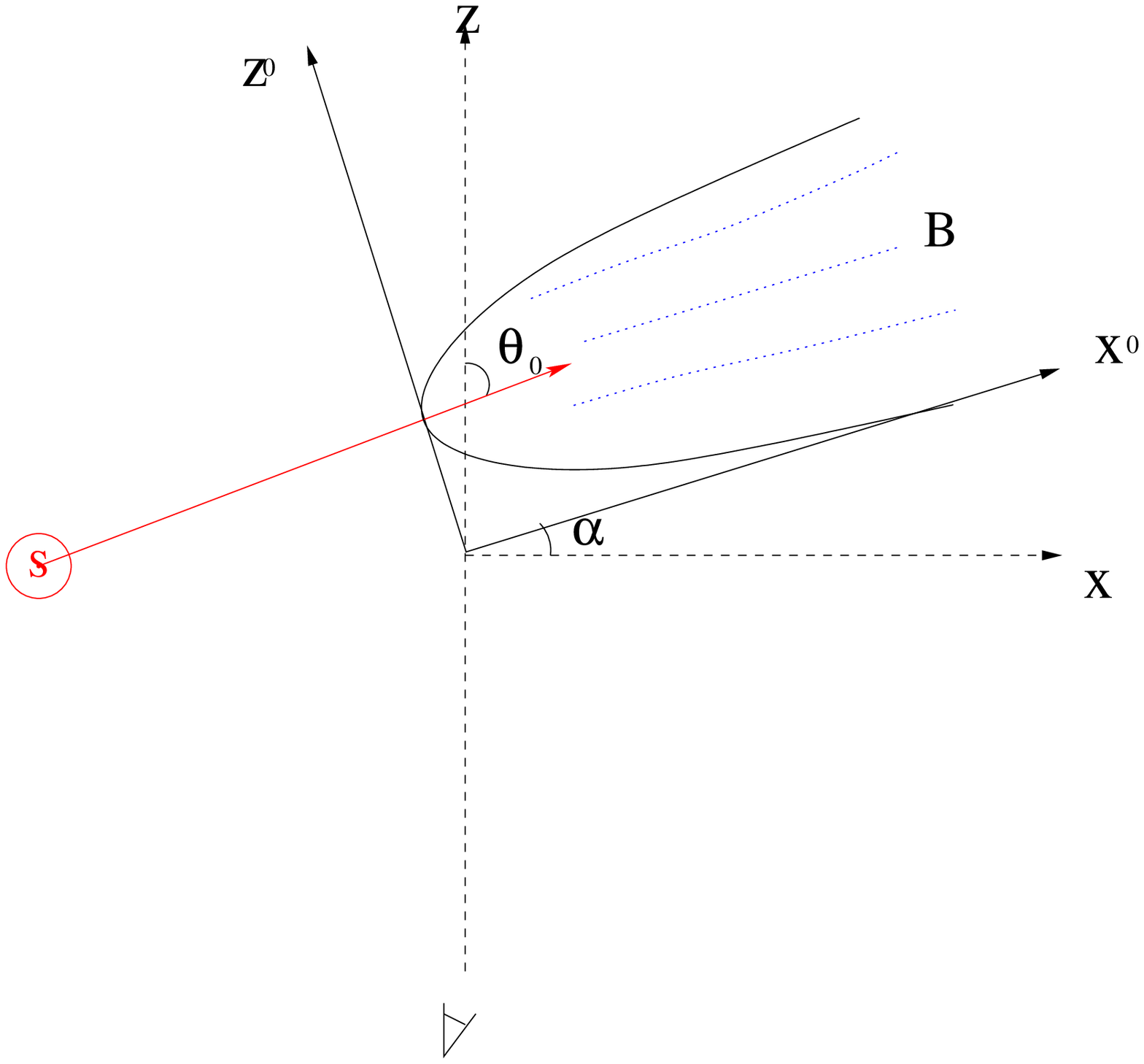}
\hfil
\includegraphics[width=.4\textwidth]{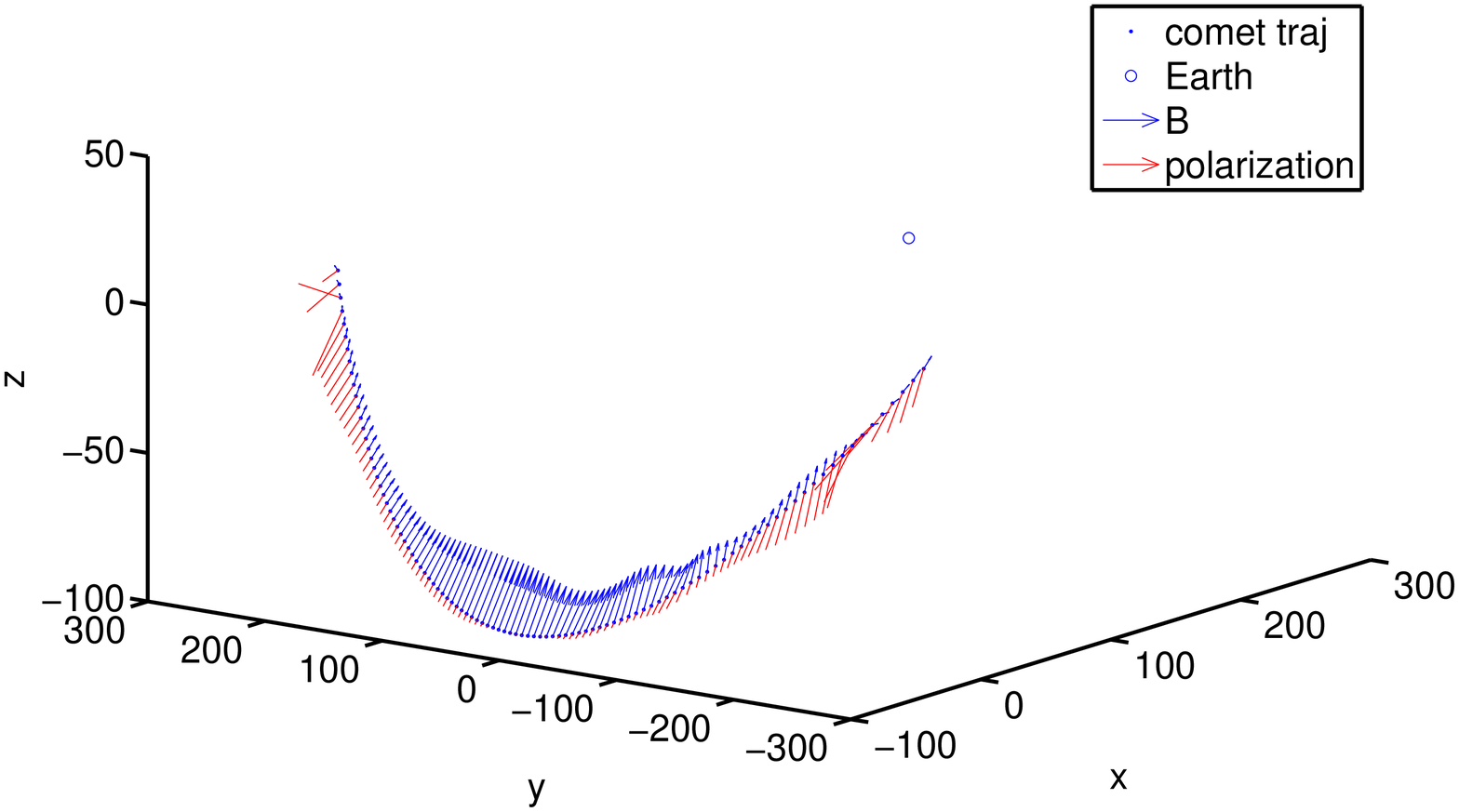}
}
\caption{{\it Left Panel}. Schematic of the resonance scattering of sunlight by the sodium in comet wake. The sodium tail is 
points in the direction opposite to the Sun. The observer on the Earth sees the stream at the angle $\theta_0$. Magnetic field
realigns atoms via fast Larmor precession. Thus the polarization traces the interplanetary magnetic fields.  {\it Right Panel}. The magnetic field from the model of inner heliosphere (blue lines) (Toth et al. 2005) and the predicted vectors of linear polarization of Sodium D2 emission (red lines) along the comet trajactory. From Yan et al. 2008.}
\label{cometmag}
\end{figure*}

The structure of the magnetic field in the heliosphere can be studied by the polarization of Sodium D2 emission in the comet's wake. Though the abundance of sodium in comets is very low, its high efficiency in
 scattering sunlight makes it a good tracer (Thomas 1992). As discussed in Yan \& Lazarian (2007), the gaseous sodium atoms in the comet's tail acquire angular momentum from the solar radiation, i.e. they are aligned. Resonant scattering from these aligned atoms is polarized. Distant from comets, the Sun can be 
considered a point source. As shown in Fig. \ref{cometmag}, the geometry of the 
scattering is well defined, i.e., the scattering angle $\theta_0$ is known. The alignment is modulated by the local magnetic field. The polarization of the sodium emission thus provides exclusive 
information on the magnetic field in the interplanetary medium. We take the data cube from the spacecraft measurement as described above. Depending on its 
direction, the embedded magnetic field alters the degree of 
alignment and therefore the polarization of the light scattered by the aligned atoms. Fig.\ref{cometmag} illustrates the trajectory of a comet along which the magnetic field varies and the polarization of Sodium D2 emission changes accordingly. By comparing observations with it, we can determine the structure of magnetic field in the heliosphere. For interplanetary studies,
one can investigate not only spatial, but also temporal variations
of magnetic fields. Since alignment happens at a time scale $\tau_R$, magnetic field variations on this time scale will be reflected. This can allow for a cost-effective way of studying
interplanetary magnetic turbulence at different scales.
On the basis of the results above we expect that comets may become an important source of information about interplanetary magnetic fields and their variations.

\section{Discussion and Summary}

The goals of the review above were, first of all, to demonstrate the necessity of detailed studies of magnetic turbulence for modeling of various
astrophysical environments, in particular, the environment of the Local Bubble and, secondly, to appeal for the necessity of new ways of studying magnetic fields in solar wind. To address the former goal we discussed the properties of MHD turbulence in different environments (fully ionized and partially ionized gas) for different ways of excitation (balanced and imbalanced turbulence). The second goal was addressed by considering a new promising way of tracing magnetic field structure in the interplanetary medium. 

The examples are representative of authors' interests, but they vividly show that the "brute force approach" to simulating Local Bubble and other astrophysical environments may easily fail. For instance, we showed that it would not be fruitful for calculating of the intermittency of environmental heating and it is likely to fail in representing the effects of subdiffusion. It also  may deliver unreliable results when simulating heat transfer in magnetized plasmas etc. Therefore detailed quantitative modeling may require, first of all, the creation of the "tool box" of the particular recepiees of how to parameterize particular properties of turbulent fluid. We claimed that studies of these properties were not only desirable, but were absolutely necessary. For instance, unless we reach a consensus on the rate of reconnection in collisional environments, the simulations of the interstellar gas will be highly suspect (see the corresponding discussion in \S 5.2). 
All in all, we believe that the progress can be achieved via better understanding of fundamental properties of magnetized plasma and aggressive testing the results of modeling with observations. The fact that the properties of turbulent fluid are important for a wide range of astrophysical problems makes it easier to concentrate efforts and resources on such studies. This also calls for more vigorous scientific exchanges between the disciplines.

Our points of the review above can be briefly summarized in the following way:\\
1. Properties of astrophysical fluids, including those of solar wind, Local Bubble and interstellar turbulence are, to large extend, determined by magnetic turbulence. The "brute force" approach aimed at detailed modeling of the aforementioned astrophysical environments keeping magnetic turbulence realistic is doomed. Instead, we feel that focused studies of particular physical phenomena in turbulent magnetized plasma can clarify when the simulations that take into account many processes at once provide a correct physical picture. These studies can help parameterize effects of turbulence in numerical codes.\\
2. By considering imbalanced turbulence, as this is the case of the Solar wind turbulence, and high-$Pr$ number turbulence, that can
approximate the turbulence in the partially ionized plasma, we demonstrated that magnetic turbulence can have many different properties. These
properties affect the transport properties of the magnetized media, e.g. scattering of cosmic rays, the formation of density enhancements, including the formation of small ionized and neutral structures (SINS), the change of magnetic topology, i.e. magnetic reconnection, thermal conductivity.\\
3. We advocate the approach to simulating processes in the aforementioned environments that uses the synergy of focused studies of particular processes in turbulent magnetized plasmas, numerical simulations of astrophysical situations and observational studies. We also feel that collecting of new data on magnetic fields is essential. In the current situation, it is promising to explore new techniques of magnetic field studies, e.g. the ones that makes use of atoms aligned in their ground state by radiation and realigned by the external magnetic field. In particular, we discuss possible studies of interplanetary turbulence using Na aligned atoms in a comet wake.


{\bf Acknowledgements}   AL research is supported by
by the NSF Center for Magnetic Self Organization in
Laboratory and Astrophysical Plasmas and NSF Grant AST-0808118. MO research is supported by NSF CAREER grant 201467. 


\end{document}